\documentclass[11pt,twoside]{article}

\usepackage{dsfont}
\usepackage{graphicx, mwe}
\usepackage{subcaption}
\usepackage{amsmath}
\usepackage{amssymb}
\usepackage{enumitem}
\usepackage{algorithm}
\usepackage{algorithmicx}
\usepackage{algpseudocode}
\usepackage[page,titletoc,toc,title]{appendix}
\usepackage{amsthm}
\usepackage{titling}
\usepackage{tabularx}
\usepackage{longtable}
\usepackage{multirow}
\usepackage{caption}
\usepackage{changepage}
\usepackage{hyperref}
\usepackage[a4paper,hmargin=2.8cm,vmargin=2.0cm,includeheadfoot]{geometry}
\usepackage{authblk}

\title{Smoothness and other hyperparameter estimation for inverse problems related to data assimilation}
\author[1]{Baptiste Simandoux}
\author[1]{Nikolas Kantas}
\author[1]{Dan Crisan}
\affil[1]{Department of Mathematics, Imperial College, London, SW7 2AZ, UK.}
\date{}
\begin{document}
	
	\maketitle	
	
	\begin{abstract}
		We consider Bayesian inverse problems arising in data assimilation for dynamical systems governed by partial and stochastic partial differential equations. The space–time dependent field is inferred jointly with static parameters of the prior and likelihood densities.  
		Particular emphasis is placed on the hyperparameter controlling the prior smoothness and regularity, which is critical in ensuring well-posedness, shaping posterior structure, and determining predictive uncertainty. Commonly it is assumed to be known and fixed a priori; however in this paper we will adopt a hierarchical Bayesian framework in which smoothness and other hyperparameters are treated as unknown and assigned hyperpriors. Posterior inference is performed using Metropolis-within-Gibbs sampling suitable to high dimensions, for which hyperparameter estimation involves little computational overhead. 
		The methodology is demonstrated on inverse problems for the Navier–Stokes equations and the stochastic advection–diffusion equation, under sparse and dense observation regimes, using Gaussian priors with different covariance structure. Numerical results show that jointly estimating the smoothness substantially reduces the errors in uncertainty quantification and parameter estimation induced by smoothness misspecification, by achieving performance comparable to scenarios in which the true smoothness is known.
	\end{abstract}
	
	\section{Introduction}
	
	In meteorology, oceanography and atmospheric sciences, hydrology, geophysics, various dynamical space-time dependent quantities need to be estimated and used for prediction; see \cite{carrassi2022data, ghil_data_1991, bocquet_beyond_2010, bennett_inverse_2002} for an overview. They are typically modelled with physical models from fluid dynamics formulated as partial differential equations (PDEs) and more recently stochastic PDEs (SPDEs), describing the evolution of the atmospheric quantities such as velocity, temperature, pressure, turbulence of wind or ocean currents, salinity and many others \cite{wikle_spatiotemporal_2001, beletsky_modeling_2003, majda_filtering_2012}. In the last few decades, increasingly larger space–time data sets are obtained from remote sensing satellites or in situ sensors for example \cite{hey_data_2003}. Satellites produce high-resolution images of geophysical variables, such as stratospheric ozone, sea surface height winds, that are used for many tasks ranging from the estimation of seasonal cycles to the identification of long-term trends \cite{stroud_ensemble_2010, stein_spatial_2007}. 
	Data assimilation provides a framework to combine this observational data with dynamical models to estimate and predict accurately those states. Forecasting, one of the core aims of numerical weather prediction, can be achieved using an inferred initial condition of the state by propagating its governing dynamics forward. As these model trajectories quickly deviate from their true state, we incorporate observations to correct this drift, see \cite{reich_probabilistic_2015, bennett_inverse_2002, carrassi2022data} for more details. In addition to states, referred to as dynamical variables, there are static parameters that determine the model equations and properties such as smoothness, noise amplitude, rotation angles among others. These also need to be estimated from data and in this paper we will consider joint Bayesian inference for dynamic and static variables \cite{agapiou_analysis_2014, chen_dimension-robust_2019}. 
	
	Data assimilation problems are naturally phrased as Bayesian inverse problems: this has been common notably in the fields of meteorology, atmospheric and oceanic sciences, see \cite{carrassi2022data, ghil_data_1991} for examples. Motivated by these applications, we adopt a Bayesian perspective to infer from noisy observations a vector field $v=\left(v(t,x); t\in[0,T],\: x\in\mathcal{X}\right)$ in a suitable separable vector space  $(V, \|\cdot\|_{V})$. 
	Bayesian inverse problems have grown significantly in popularity over the last decade, driven by the availability of simulation-based algorithms and large computational power \cite{kaipio_statistical_2005}. Within this framework, we aim to learn a posterior distribution $\mu$ given data $y$ via the abstract Bayes rule:  
	\begin{equation}\label{bayes}
		\frac{d\mu}{d\mu^\theta_0}(v)=l(y,v,\phi).
	\end{equation}
	Here $\theta, \phi$ denote sets of various constants used to specify the prior $\mu_0$ and the likelihood $l$ respectively. An estimate of $v$ is given by samples of this posterior. The prior acts as regulariser and imposes certain smoothness and regularity conditions on $v$ so that $\mu$ is well posed  \cite{engl1996neubauer, cotter_approximation_2010, konen_data_2025, stuart_inverse_2010}. Different regularisers can be chosen depending on the approach used, for example Gaussian priors correspond to a Tikhonov regulariser \cite{hein_tikhonov_2009} or a Total Variation prior can be used to preserve edges \cite{lassas_can_2004}. Here we will use Gaussian priors as in \cite{knapik_bayesian_2012} but our methodology is generic for other priors and regularisers, see \cite{engl1996neubauer} for an overview of different options.
	
	$\theta$ is usually a concatenation of several parameters: variance amplitude, spatial range, correlation length or order of a differential operator for example. In this paper we will pay particular attention to determining the smoothness/roughness of $v$, denoting the relevant parameter as $\alpha$. One requirement for the a priori spatial smoothness $\alpha$ is to ensure posterior well-posedness of the inverse problem itself. Beyond this $\alpha$ is crucial in  setting various spatial properties of the posterior and subsequent predictions. 
	In many cases in data assimilation estimation of $\alpha$ is omitted and is assumed to be fixed and known \cite{emzir_non-stationary_2020, kantas_sequential_2014, pons_llopis_particle_2018, roininen_hyperpriors_2018, sigrist_stochastic_2015}. 
	Similar challenges in setting these regularisation parameters also appear in image processing problems, see \cite{vidal_maximum_2020, molina_bayesian_1999}. 
	In this paper we propose to estimate $\alpha$ and demonstrate the benefits of the added model flexibility brought by the hyperpriors. The remaining parameters of $\theta$ and the ones of $\phi$ will also need to be estimated. While in some cases these may be known a priori via some offline calibration methods, especially for $\phi$, they often require joint estimation with $v$, particularly when not physically measurable \cite{agapiou_analysis_2014, coullon_markov_2022}.
	
	To that end, we will adopt a hierarchical modeling approach \cite{agapiou_analysis_2014, chen_dimension-robust_2019, gelman_bayesian_2015, kim_hierarchical_2023, papaspiliopoulos_non-centered_2003, papaspiliopoulos_general_2007}, treating $(\theta,\phi)$ as variables also to be inferred and assigned with hyper-prior $\pi(\phi,\theta)$. This allows for a flexible and expressive way to specify priors and models. The resulting joint prior distribution on $v,\theta,\phi$ is chosen to lead to a well-posed posterior distribution by conditioning on the observed data 
	\cite{stuart_inverse_2010}. 
	It is common in hierarchical inverse problems for $v$ to employ ensemble Kalman techniques \cite{kim_hierarchical_2023, sanz2025hierarchical}. 
	Whilst fast and cheap to implement, the latter are not theoretically justified and are biased. They have performed well in providing point estimates like the \textit{maximum a posteriori} (MAP) estimator, however they do not capture the uncertainty around these estimates \cite{agrawal_variational_2022}. 
	This motivates our choice to use Monte Carlo methods for parameter estimation and in particular Markov Chain Monte Carlo (MCMC) as in \cite{chen_dimension-robust_2019, cotter_bayesian_2009, cotter_approximation_2010, cotter_mcmc_2012, law_proposals_2014, luengo_survey_2020, neal_monte_1997, stuart_inverse_2010}. To estimate $\phi, \theta$ along with $v$, we will use Metropolis-within-Gibbs (MwG) sampling. For $v$, we will use appropriate MCMC algorithms that are efficient and robust to the high dimensionality of $v$, namely the preconditioned Crank–Nicolson (pCN) \cite{cotter_mcmc_2012, hairer_spectral_2014, law_proposals_2014}. 
	These are computationally intensive methods especially when used for numerical discretisations of $v$ and the relevant (S)PDEs, but they are theoretically justified and robust under mesh refinement \cite{hairer_spectral_2014}.
	
	We will demonstrate the merit of our approach in two case studies. First, we will consider an inverse problem using the Navier-Stokes equations,  
	similar to \cite{cotter_bayesian_2009, cotter_approximation_2010, kantas_sequential_2014, stuart_inverse_2010}.  
	Secondly we consider the stochastic advection-diffusion equation, popular for environmental monitoring, for example in \cite{stroud_ensemble_2010, xu_estimation_2007, sigrist_stochastic_2015, duan_modeling_2009, malmberg_interpolating_2008, zheng_hierarchical_2010}. 
	For $\mu_0$, we will use the popular Gaussian process prior used in many data assimilation settings \cite{stuart_inverse_2010, kaipio_statistical_2005, knapik_bayesian_2012, nickl_posterior_2024, kantas_sequential_2014}. 
	For the Navier-Stokes case, the covariance of $\mu_0$ will be set as $\beta^2A^{-\alpha}$, where $\alpha$ and $\beta$ characterise the smoothness and scaling respectively and A is the Stokes operator. 
	In the stochastic advection-diffusion case we consider as prior the Whittle-Matérn covariance \cite{whittle_stationary_1954}, widely used in spatial statistics \cite{monard_statistical_2021, sigrist_stochastic_2015}. In both cases the posterior is highly sensitive to choices of its hyperparameters \cite{neal_monte_1997, nelsen_bilevel_2025} and in particular $\alpha$.
	
	\subsection{Relations with existing literature}
	
	From a theoretical standpoint a fixed $\alpha$ may yield sub-optimal contraction rates, unless the prior captures the fine properties of the truth like its regularity \cite{knapik_bayesian_2012}. As this is rarely known, determining the regularity/smoothness of $v$ systematically is necessary for accurate inference of $v$. 
	Alternatively to Hierarchical Bayesian methods, one could use empirical Bayes to estimate $\theta$, including $\alpha$ from the data, typically by maximizing the marginal likelihood of $\theta$ prior to posterior inference \cite{teckentrup_convergence_2020, knapik_bayes_2016}. 
	Whilst computationally cheaper, this is not a fully Bayesian method and it ignores the parameters' uncertainty, potentially underestimating our posterior uncertainty. 
	Hierarchical Bayesian methods considering smoothness were studied theoretically for the linear diagonalizable inverse problems in \cite{knapik_bayes_2016}, obtaining adaptive optimal convergence rates, under mild assumptions on $\alpha$'s hyper-prior. The specific form of the parameterisation for $\theta$ can affect the mixing of MwG \cite{agapiou_analysis_2014, papaspiliopoulos_non-centered_2003, yu_center_2011}. Non-centered parameterisations of the latent space and hyperparameters allow for faster mixing in MwG, and a non-centered parameterisation was also used for pCN algorithms, with \cite{agapiou_analysis_2014} showing that it is robust to discretisation. 
	We note that to the best of our knowledge, there are no case studies related to nonlinear inverse problems prior to this work investigating and demonstrating the efficiency of this approach when also estimating the smoothness parameter $\alpha$.
	
	We will be using the MwG algorithm along with pCN as in \cite{chen_dimension-robust_2019}. 
	Parameter learning for the Matern prior using MwG is studied in \cite{roininen_hyperpriors_2018} but $\alpha:=3-d/2$ was fixed. 
	For linear models the MAP estimate has been well studied theoretically in \cite{knapik_bayes_2016, dunlop_hyperparameter_2020}, but there is a lack of case studies on inverse problems with PDEs/SPDEs, especially when learning the smoothness parameter. Our paper addresses this gap and is intended to illustrate that one can recreate similar performance to knowing $\alpha$. In parallel work, \cite{nelsen_bilevel_2025} provides a case study with a point estimate (MAP) for fixed hyperparameters, without considering likelihood parameters $\phi$. We will also demonstrate how learning $\alpha$ leads to better uncertainty quantification. Finally our case studies include different models and observation regimes. 
	
	\subsection{Contributions}
	\begin{itemize}
		\item We propose an efficient MwG algorithm for inverse problems of PDEs and SPDEs. We use pCN for the state and Gibbs sampling or Gibbs MCMC kernels for the hyperparameters, as in \cite{chen_dimension-robust_2019}. We remark that the computational overhead of considering hyperparameters is minimal.
		\item We pay particular attention to the effect of the smoothness parameter on the state's inference and prediction, often overlooked in the literature as mentioned earlier. To that end, we demonstrate its improvements in uncertainty quantification both in estimation and prediction compared to standard methods relying on fixed smoothness assumptions. We also look into the smoothness' influence on the rest of the parameters' estimation.
	\end{itemize}
	
	\subsection{Organisation}
	
	The structure of this paper is as follows: in section~\ref{sec_problem} we present the problem formulation, our notation along with the two different problems to be studied. Section~\ref{sec_MwG} describes the Metropolis-within-Gibbs algorithm we are considering in the high-dimensional setting and its implementation for each case study. Subsequently, section~\ref{sec_study} contains numerical results. Finally, in section~\ref{sec_conclusion} we will conclude and discuss future work. The Appendix contains supplementary material and additional numerical experiments.
	
	\section{Problem Formulation}\label{sec_problem}
	
	Adopting a Bayesian perspective, we will estimate $v$ from $y$ by sampling from its posterior distribution, introduced in \eqref{bayes}. To do so, we first specify a prior for $v$ conditional on parameter $\theta$, where $\alpha\in\theta$ controls the spatial regularity of the prior realizations. We will model the observations \( y \) as noisy measurements of the latent field \( v \) at fixed locations and discrete intervals for the observation times. As a result, the likelihood function will depend on the dynamics of $v$ and the observation process.  For simplicity we will only consider an Eulerian observation setting with a regularly spaced grid for the observation locations and use Gaussian additive noise, but we point out that our methodology straightforwardly extends to more general non-linear non Gaussian observation schemes and irregular grids as we are using MCMC.
	
	Recall that $\phi, \theta$ are the parameters of the likelihood and prior respectively. As mentioned above, inference for $\theta$ and $\phi$ is performed jointly with $v$. Typically, the parameters $\theta, \phi$ are low dimensional whilst the dimension of $v$ is high, so the latter dictates the algorithm's computational cost. Below we will introduce the observation equation and two different dynamical models we will consider in our numerical study: the 2D Navier-Stokes equation and the stochastic advection-diffusion equation.
	
	\subsection{Observations}\label{sec_obs}
	
	We model the data as noisy measurements of the velocity field $v$ on a fixed grid of points: $x_{1},...,x_{\Upsilon,}$ for $\Upsilon\geq 1$.
	We assume that the observation noise is Gaussian and its standard deviation $\tau$ is known. These are obtained regularly at intervals $\delta$ time units apart.
	$$
	y_{n,\upsilon} = v(x_{\upsilon} ,n \delta)+\tau \zeta_{n,\upsilon},\quad \zeta_{n,\upsilon} \overset{iid}{\sim} \mathcal{N}(0, 1) ,\quad 1 \leq \upsilon \leq \Upsilon, 1 \leq n \leq T.
	$$
	So the likelihood of this data conditioned on the initial condition $v_0$ takes the form:
	
	\begin{equation}\label{likeli eq}
		l(y; v)= \frac{1}{(2\pi\tau^2)^{\frac{T\cdot\Upsilon}{2}}} \prod\limits_{n=1}^{T} \prod\limits_{\upsilon=1}^{\Upsilon} \exp\left(-\frac{1}{2\tau^{2}} ( y_{n,\upsilon} -  v(x_{\upsilon} ,n \delta))^{2}\right).
	\end{equation}
	
	\subsection{Case study 1: Inferring the initial condition of deterministic Navier-Stokes equations}\label{NS_setting}
	
	The domain we consider is a 2D Torus $\mathbb{T}=[0,2\pi]^{2}$ with initial flow $v_0\in V$ and let $v : \mathbb{T} \times [0, \infty) \rightarrow \mathbb{R}^{2}, \,v(x, t)=(v_{1}(x, t),v_{2}(x, t))^T$. One can write the incompressible Navier-Stokes PDE as lifted in a suitable Hilbert space (see section \ref{NS appendix} in the Appendix for details)
	\begin{equation}\label{NS_ODE}
		\frac{dv}{dt} + \eta Av + B(v, v)=P (f ) ,\quad v(0,\cdot) = v_0,
	\end{equation}
	where $\eta>0$ is a known viscosity parameter, $A$ is the Stokes operator, $B$ is the symmetric bilinear form and $P(f)$ is the forcing term. Let ${\Psi(\cdot,t):V \rightarrow V }_{t\geq0}$ denote the semigroup of solution operators for the ODE \eqref{NS_ODE} through t time units. Due to the nature of the observations we are interested in the discrete-time semigroup where $ v(\cdot ,n \delta)=G^{(n)}_{\delta}(\cdot)=\Psi(\cdot,n \delta)$ corresponding to time instances $t = n \delta$, of lag $\delta>0$ and $n =0,...,T$, with conventions $G_\delta^{(0)} =I,\; G_\delta^{(1)} = G_\delta, \; G_\delta^{(n)}=G_\delta * G_\delta^{(n-1)}$.  For simplicity we will assume $\phi=\{\eta, P(f), \tau\}$ to be known. Notice then that in this problem the Navier-Stokes dynamics enter in the posterior only in the likelihood function, formulated above in \eqref{likeli eq}. Also all the information for $v$ is captured in $v_0$ as $v(x_{\upsilon} ,n \delta)=G^{(n)}_{\delta}(v_0)(x_{\upsilon})$ so we will consider this as an inference problem for $v_0$. To clarify, since the dynamics are deterministic, forecasting can be performed by propagating them from samples of $v_0$ to the desired future time.
	
	An appropriate orthonormal basis for $V$ is comprised of 
	$$\psi_{k}(x)= \frac{k^{\bot}}{2\pi|k|} \exp (ik \cdot x), k \in Z^{2}\setminus\{0\}\text{, where }k^{\bot} =(-k_2,k_1)^T.$$
	Using this basis we  obtain the Fourier decomposition where $v_0(x)=\sum\limits_{k\in \mathbb{Z}^{2}\setminus{\{0\}}} u_k \psi_k(x)$ with $u_k=\langle {v_0}, {\psi_k} \rangle$. Since $v_0$ is real valued, $\overline{u_{k}} = -u_{-k}$ so we only need to consider half the Fourier coefficients. We split the domain by defining:
	$$
	\mathbb{Z}^{2}_{\uparrow} = \{k = (k_{1},k_{2}) \in \mathbb{Z}^{2}\setminus\{0\}: k_{1} + k_{2} > 0\}\cup\{k \in \mathbb{Z}^{2}\setminus\{0\}: k_{1} + k_{2} =0,k_{1} > 0\}\;,
	$$
	and imposing $\overline{u_{k}} = -u_{-k}$ for $k\in \{\mathbb{Z}^{2}\setminus\{0\}\} \setminus Z^{2}_\uparrow$. We represent the initial vector field $v_0$ by its Fourier coefficients, and equivalently:
	$v_0\leftrightarrow\{u_k\}_{k\in\mathbb{Z}^2_\uparrow}$. $v_0$ will denote the concatenated vector of the real and imaginary part of its Fourier coefficients.
	
	Since we are only inferring $v_0$, we set the prior on $v_0$ to be $\mu_{0}=\mathcal{N}(0, \beta^{2}A^{-\alpha})$ with hyperparameters $\theta:= (\alpha>1/2, \beta>0)$ as earlier. This prior, a Gaussian distribution on this Hilbert space, is a convenient and flexible prior, see for details \cite{da_prato_stochastic_2014}. In the Fourier domain we have:
	$$v_0\sim \mu_{0}\Longleftrightarrow \text{Re}(u_{k}),\, \text{Im}(u_{k})\overset{iid}{\sim}\mathcal{N}(0, \frac{1}{2}\beta^{2}|k|^{-2\alpha}), \;k\in \mathbb{Z}^{2}_\uparrow\,.$$	
	The prior admits a Karhunen–Loève expansion: \begin{equation}\label{prior_KL_exp.}
		\mu_{0}^\theta=Law\left(\sum\limits_{k\in \mathbb{Z}^{2}\setminus\{0\}} \frac{\beta}{\sqrt{2}} |k|^{-\alpha}\xi_{k}\psi_{k}\right),
	\end{equation}
	with $\theta=(\alpha,\beta^2)$ , $\text{Re}(\xi_{k}), \text{Im}(\xi_{k})\overset{iid}{\sim}\mathcal{N}(0,1),\, k\in \mathbb{Z}^{2}_{\uparrow};\;\;\xi_{k}=-\overline{\xi_{-k}}, \,k\in\{\mathbb{Z}^{2}\setminus\{0\}\}\setminus \mathbb{Z}^{2}_{\uparrow}$, providing the Non-Centered Parameterisation $(\theta, \{\xi_k\}_{k\in \mathbb{Z}^{2}_{\uparrow}})$.
	
	We choose a zero-mean prior for simplicity; results will still hold for any mean function in the Cameron-Martin space of $\mu_{0}$ \cite{cotter_bayesian_2009, cotter_approximation_2010, stuart_inverse_2010}. Under this prior, $u_k$ are initially assumed independent normally distributed, with a particular rate of decay for their variances as $|k|$ increases so that the prior is trace class i.e. $$\sum\limits_{k\in \mathbb{Z}^{2}\setminus\{0\}} \frac{\beta}{\sqrt{2}} |k|^{-\alpha}<\infty.$$ This variance decay is particularly relevant for the Navier-Stokes model, for which the dissipation will transfer energy from high to low frequencies until $v_0$ enters the PDE's attractor \cite{robinson_infinite-dimensional_2003}. For appropriate forcing and initialisation, the spectrum of $v_0$ tends to concentrate at low frequencies. This means higher frequencies will be less informative in the posterior. Here the role of the smoothness parameter $\alpha$ is crucial: higher values would correspond to lower number of Fourier modes capturing most of the prior and posterior and since the viscosity constant $\eta$ is fixed and known, $\alpha$ sets the effective dimension.
	
	Since $\phi$ is known, we use the following joint posterior probability measure $\mu$ on $V$:
	\begin{equation}\label{posterior}
		\mu(v_0,\alpha,\beta^2)= \frac{1}{Z(y)}l(y;v_0)   \mu^{\alpha, \beta^{2}}_{0}(v_0)\pi(\alpha)\pi(\beta^2),
	\end{equation}
	where the superscript in $\mu_0^{\alpha, \beta^{2}}$ denotes conditioning with $\alpha, \beta^{2}$ and $\pi(\alpha),\pi(\beta^2)$ are independent hyper-priors.
	The well-posedness of $\mu$ is covered by the results in \cite{cotter_bayesian_2009, cotter_approximation_2010, stuart_inverse_2010}, which deal with the infinite dimensional variable $v_0$, and has been established with the requirement that $\alpha>1/2$ or equivalently $A^{-\alpha}$ be a trace-class operator. As long as the hyper-prior $\pi(\alpha)$ does not violate this requirement and $Z(y)$ is finite, the posterior will be well posed. For this reason we will use an inverse Gamma distribution for $\beta^2$ and an appropriate uniform for $\alpha$. 
	
	\subsection{Case study 2: Data assimilation with the Stochastic Advection Diffusion equation}\label{SAD_setting}
	
	We now study a model whereby the state's dynamics follow a linear SPDE on $V:=\mathbb{T}^2$.
	
	$$ \text{d}v(t) = (-\mu^{T} \nabla v_{t} + \nabla\cdot \Sigma \nabla v_{t} - \zeta v_{t})\text{d}t + \text{d}\epsilon_{t},$$
	where $\mu^{T} \nabla v_{t}$ models transport effects, advection in weather applications, with drift $\mu$, $\nabla\cdot \Sigma \nabla v_{t}$ is a diffusion term that can incorporate anisotropy and $- \zeta v_{t}, \zeta > 0$ is a damping term. The stochastic noise $\epsilon_{t}$ is a temporally white and spatially correlated Gaussian process, with Whittle-Matérn covariance.
	
	$$\Sigma^{-1}= \frac{1}{\rho_1^2}\begin{bmatrix} \cos(\psi)  & \sin(\psi) \\ 
		-\gamma \sin(\psi)  & \gamma\cos(\psi)\end{bmatrix}^{T}\begin{bmatrix} \cos(\psi)  & \sin(\psi) \\ 
		-\gamma \sin(\psi)  & \gamma\cos(\psi)\end{bmatrix}$$
	where $\rho_{1} > 0, \gamma > 0$ and $\psi \in [0, \pi/2]$. Here, $\rho_{1}$ acts as a range parameter and controls the amount of diffusion. $\gamma$ and $\psi$ control the amount and the direction of anisotropy; with $\gamma = 1$, isotropic diffusion is obtained. For this problem: $\phi=(\zeta,\rho_1, \gamma,\psi,\mu_1,\mu_2, \tau)$. Contrary to the previous setting, the noise in the dynamics means we cannot easily recover $v_t$ directly from $v_0$. To do this we need to simulate the whole path of $v_t$. As a result, this is an inverse problem on the whole path $\{v_t\}_{t\in[0,T]}$. Forecasting beyond the time of the last observation $T$ is possible after inferring $\{v_t\}_{t\geq0}^{T}$ given observations defined as in section~2.1. This filtering setting and the prior are defined on the path space $\{v_t\}_{t\geq0}^{T}$. So we let our prior include the initial condition and the noise increments in the dynamics, which will depend on $\theta$. We set $v_0\overset{\mathbb{P}}{:=}\epsilon_0$ and describe in detail the distribution of $\epsilon$ below.
	
	\subsubsection{Whittle-Matérn prior}
	
	The Matérn covariance between locations $u,v$ is 
	$$r(u, v) = \frac{\sigma^{2}}{\Gamma(\nu)2^{\nu-1}} (\rho_0\|v - u\|)^{\nu} K_{\nu}(\rho_0\|v - u\|),$$
	where $K_{\nu}$ is the modified Bessel function of second kind and order $\nu > 0, \rho_0 > 0$ is a scaling parameter  and $\sigma^{2}$ is the marginal variance. Using a Gaussian field with Whittle-Matérn covariance, the noise $\{\epsilon_t\}_{t=0}^T$ is quite a common choice in practice and applications \cite{monard_statistical_2021, sigrist_stochastic_2015, stroud_ensemble_2010, malmberg_interpolating_2008}.
	
	A Gaussian field $x(u)$ with Matérn covariance is a solution to
	\begin{equation*}\label{matern eq}
		(\rho_0^{2} - \Delta)^{\alpha/2}x(u) = \sigma\,\mathcal{W}(u),\quad u\in \mathbb{R}^d,\quad \alpha = \nu + d/2,\;\; \rho_0 > 0,\; \nu > 0
	\end{equation*}
	where the innovation process $\mathcal{W}$ is a spatial Gaussian white noise \cite{lindgren_explicit_2011}. 
	As in \cite{sigrist_stochastic_2015} we will infer $\sigma^2$, however in our case we will also infer $\alpha$ (\cite{sigrist_stochastic_2015} has it set with $\nu=1$). 
	$(\rho_0^{2} - \Delta)^{\alpha/2}$ is a pseudo-differential operator with Fourier transform:
	\begin{equation}\label{matern op}
		\{\mathcal{F}(\rho_0^{2} -\Delta)^{\alpha/2}\Lambda\}(k) = (\rho_0^{2} + \|k\|^{2})^{\frac{\alpha}{2}}(\mathcal{F}\Lambda)(k)
	\end{equation}
	where $\Lambda$ is a function on $\mathbb{R}^{d}$ for which the right-hand side of \eqref{matern op} has a well-defined inverse Fourier transform. $\rho$ controls the spatial correlation, a defining feature of the Matérn field. Here $\theta=(\alpha, \sigma^2, \rho_0)$, so $(\rho_0^{2} + \|k\|^{2})^{\frac{\alpha}{2}}$ cannot be precomputed, and we consider independent hyperpriors for $\theta$. As a result in this case every model parameter is being estimated. So the  joint posterior we sample from corresponds to:
	\begin{equation}\label{posterior_SAD}
		\mu(v,\phi,\theta)= \frac{1}{Z(y)}l^\phi(y;v)   \mu^{\theta}_{0}(v)\pi(\theta)\pi(\phi).
	\end{equation}
	
	\section{Monte Carlo Methods for filtering and inverse problems}\label{sec_MwG}
	
	\subsection{Markov Chain Monte Carlo}\label{MCMC}
	
	MCMC is an iterative procedure simulating a long run of an ergodic time-homogeneous $\mu$-invariant Markov chain to sample from the target distribution $\mu$. After a burn-in period the samples of the chain form an empirical approximation of $\mu$. Several Markov chain methods are available to sample from a given target posterior distribution $\mu$, such as the Gibbs sampler or the Metropolis-Hastings algorithm; see \cite{tierney_markov_1994} for an overview. 
	
	Our setting requires sampling from distributions defined on a Hilbert space and the posteriors lie in an infinite-dimensional state-space. For inverse problems on $v$, one cannot easily use  conditional dependence to develop Gibbs-type samplers to update blocks of coefficients and improve efficiency. 
	In practice we use a high-dimensional discretisation, but our method still needs to be robust to mesh refinement. 
	Sampling directly from the full conditional of $v$ given $\phi, \theta, y$ is impossible so we use a MCMC suited to high dimensions. Random-Walk Metropolis algorithms are not a valid option for targets with Gaussian priors as their spectral gaps will provably collapse with increasing dimensions \cite{hairer_spectral_2014}. 
	Autoregressive proposals that are invariant to the prior have been proposed in \cite{tierney_markov_1994}. These were recently renamed as preconditioned Crank–Nicolson (pCN) and introduced for inference of high-dimensional inverse problems in \cite{cotter_mcmc_2012, stuart_inverse_2010}. \cite{hairer_spectral_2014} establishes that in contrast to Random walk methods, the spectral gap of pCN and hence the mixing of the method does not collapse as the dimension grows and overcomes some of the slow mixing issues encountered by standard MCMCs in high dimensions. We fix $\phi, \theta$ to the current estimates and run iterations of pCN on the full conditional of $v$. The pseudo-code is presented in Algorithm \ref{Alg. MCMC}.
	
	\begin{algorithm}
		
		\caption{pCN $\mathcal{K}_{pCN}(\mathbf{v_0},\;\cdot\;)$: $\mu$-invariant MCMC for High Dimensional Inverse Problems}\label{Alg. MCMC}
		
		\textit{Inputs}:
		\begin{itemize}
			\item Prior $\mu_0$
			\item step-size $\rho$ close to 1 ($\sim 1-1/(\text{number of variables to estimate})^2$)
			\item Data $y$, likelihood and its parameters $\phi$ conditioned on initial condition $v_0$: $l(y, v_0, \phi)$
			\item Number of iterations $T$ with burn-in period $burn\_in$
		\end{itemize}
		
		\textit{Initialize}: $v_0(0)\sim \mu_0$. For $t=1,\dots,T$:
		\begin{enumerate}
			\item Propose:
			$$\tilde{\mu}=\rho\cdot v_0(t-1)+\sqrt{1-\rho^2}\cdot\xi,\quad \xi\sim\mu_0$$
			\item Accept $v_0(t)=\tilde{v}_0$ with probability:
			$$1\wedge \frac{l(y,\tilde{v}_0, \phi)}{l(y,v_0(t-1), \phi)}$$
			otherwise $v_0(t)=v_0(t-1)$.
		\end{enumerate}
		
		\textit{Output:} $(v_0(t))_{t=burn\_in}^T$ form estimates of $v_0$ as samples from $\mathcal{K}_{pCN}(v_0(0))$.
		
	\end{algorithm}
	
	\subsection{Metropolis-within-Gibbs}
	
	Given the hierarchical structure of the model, we will use Metropolis-within-Gibbs (MwG). Gibbs sampling alternates sampling from the full conditional distributions of each $v, \theta, \phi$, conditional on the current values of the remaining variables and $y$. Independent hyperpriors for $\theta$ and $\phi$ will be used so that updates of $v, \phi, \theta$ will be simulating in sequence from the following densities:
	\begin{align}
		\pi(v|y, \theta, \phi) & \propto \; l(y|v, \phi)\; \pi(v| \theta)\label{conditional_v}\\
		\pi(\phi|y, v, \theta) & \propto \; l(y|v, \phi)\; \pi(v|\theta)\;\pi(\phi)\\
		\pi(\theta|v) & \propto \; \pi(v|\theta)\;\pi(\theta)\label{conditional_theta}
	\end{align}
	One could use either direct sampling from each conditional above or if this is not possible a small number of Metropolis iterations targeting each of the above densities.\footnote{One may even go further and implement MwG for each coordinate or blocks of coordinates of the variables in  $(v,\theta, \phi)$.} The latter is referred to as MwG. Sampling directly using conjugate relationships is recommended when available to accelerate sampling. 
	
	We now discuss the particular case of estimating the hyperparameters $\theta$ in the prior. Unlike $v$ and $\phi$, $\theta$ is not informed by the data, only by the estimate of $v$. So in its Gibbs update of $\pi(v, \theta, \phi|v)$ the costly likelihood does not need to be computed.
	
	\subsection{Sampling for Case study 1}\label{par. est. case 1}
	
	\begin{algorithm}[h]
		\caption{Metropolis-within-Gibbs sampler $\mathcal{K}_{MwG}(\mathbf{v_0}, \alpha, \beta^2, \;\cdot\;)$ for Case study 1}\label{Alg. parestalg}
		
		\textit{Additional inputs to Algorithm \ref{Alg. MCMC}:}
		\begin{itemize}
			\item Parameter hyper-priors: $\mathcal{P}_\beta=\text{Inv-Gamma}(a, b), \;\mathcal{U}_\alpha$
			\item proportion of steps per parameter: $p_{v_0}, \;p_{\beta^2},\; p_{\alpha}\; (=1-p_{v_0}-p_{\beta^2})$
			\item additional learning rate: $\rho_\alpha$
			
		\end{itemize}
		
		\textit{Initialize}: $v_0(0)\sim \mu_0,\;\beta^2(0)\sim\mathcal{P}_\beta,\;\alpha(0)\sim\mathcal{U}_\alpha$. For $t=1,\dots,T$:
		\begin{enumerate}
			\item Sample $p\sim U([0,1])$.
			\item IF $p<p_u$,$\;$ $v_0(t)\sim\mathcal{K}_{pCN}(v_0(t-1), \;\cdot\;|\alpha(t-1), \beta^2(t-1))$,\\
			
			ELSE IF $p<p_u+p_{\beta^2}$, 
			$\beta^2(t)\sim\text{Inv-Gamma}(a+\frac{n^2}{2}, b+ \frac{v_0(t-1)A^\alpha v_0(t-1)}{2})$,\\
			
			ELSE propose $\hat{\alpha}=\rho_\alpha\;\cdot\; \alpha(t-1) + \sqrt{1-\rho_\alpha^2} \;\cdot\; u$ where $u\sim\mathcal{U}_\alpha$, and accept $\alpha(t)=\hat\alpha$  with probability $\mathcal{A}$, or reject and set $\alpha(t)=\alpha(t-1)$.
			\item Set the remaining parameters to their previous value e.g. $ \alpha(t)=\alpha(t-1)$
		\end{enumerate}
		
		\textit{Output:} $(v_0, \alpha, \beta^2)_{t=burn\_in}^T$ form estimates of $(v_0, \alpha, \beta^2)$ as samples from $\mathcal{K}_{MwG}(v_0(0),\alpha(0), \beta^2(0), \;\cdot \;)$.
	\end{algorithm}
	
	We proceed with the Navier-Stokes inverse problem from section~\ref{NS_setting} and the observations in section \ref{sec_obs}. For the hyperparameters, $\theta=(\alpha,\beta^2)$ and the joint prior is $\pi(v| \alpha, \beta^{2})\pi(\alpha)\pi(\beta^2)$.  
	In Alg. \ref{Alg. parestalg} we present a random scan MwG where with  probability $p_v,\; p_{\beta^2}, \; p_\alpha$, we choose one of the conditionals (or when appropriate the MCMC kernels targeting them) in \eqref{conditional_v}-\eqref{conditional_theta} to iterate the corresponding variable, see step 2 in Algorithm \ref{Alg. parestalg}. 
	
	For updating $v$, the full conditional is as follows:
	\begin{align*}
		\pi(v_0|y, \alpha, \beta^{2}) & \propto \quad l(y|v_0)\; \pi(v_0| \alpha, \beta^{2})\\
		\qquad\qquad\;\;\; & \propto \quad \frac{1}{(2\pi\tau^2)^{\frac{T\cdot\Upsilon}{2}}} \prod\limits_{n=1}^{T} \prod\limits_{\upsilon=1}^{\Upsilon} \exp\left(-\frac{1}{2\tau^{2}} ( y_{n,\upsilon} - G^{(n)}_{\delta}(v_0)(x_{\upsilon}))^{2}\right)\\
		\qquad\qquad\;\;\; & \times \quad \frac{1}{\sqrt{2\pi}^{d/2}|\beta|^{d}|A^{-\alpha}|^\frac{1}{2}}\exp(-\frac{1}{2\beta^2}v_0A^\alpha v_0)\:,
	\end{align*}
	so given $\alpha, \beta^2$ we will target this density using Alg. \ref{Alg. MCMC}, where we adjust $\rho$ for an average acceptance probability ($\sim 0.2 - 0.3$). Each iteration requires running a PDE solver until the time of the latest observation to compute $l(y|v,\phi)$, which is the most computationally expensive step in our algorithm. 
	
	We will set the hyperprior $\pi(\beta^2):= \text{IG}(a,b)$ (Inverse Gamma distribution with shape parameter $a$ and scale parameter $b$) to get a conjugate full conditional:
	$$\pi (\beta^{2}|y, \alpha, v_0)\propto \pi(v_0| \alpha, \beta^{2})\pi(\beta^{2}) = \text{IG}(\hat{a}, \hat{b})$$
	with $\hat{a} = a+\frac{d}{2}, \hat{b}= b+ \frac{1}{2}v_0A^\alpha v_0$, where $d$ corresponds to the dimension of the mesh for $v$. This is equal to the number of random variables we are evaluating: $d=\frac{n(n-1)}{2}$ because of the restrictions $\nabla\cdot v =0$ and $\overline{u_k}=-u_{-k}$. Derivations for this expression can be found in the Appendix in section \ref{conj_prior}.
	
	For $\alpha$ we will use a uniform prior $\mathcal{U}_\alpha$ on compact domain and sample from pCN type MCMC kernel targeting the following full conditional:
	$$\pi (\alpha|y, \beta^2, v_0)\propto\pi(v_0| \alpha, \beta^{2})\pi(\alpha).$$
	The proposal will be $\hat{\alpha}=\rho_\alpha \alpha + (1-\rho_\alpha) \mathcal{P}$ where $\mathcal{P}\sim\mathcal{U}_\alpha$ and hence is invariant to the uniform prior. As a result, the acceptance ratio $\mathcal{A}$ for this MCMC kernel is given by:
	$$\mathcal{A} = 1 \wedge  \frac{\pi(v_0| \hat{\alpha}, \beta^{2})}{\pi(v_0| \alpha, \beta^{2})} \frac{Q(\hat{\alpha}, \alpha)}{Q( \alpha, \hat{\alpha})}= \frac{\mu_{0}(v_0|\hat{\alpha}, \beta^2)}{\mu_{0}(v_0|\alpha, \beta^2)} = \frac{|A^{-\alpha}|^\frac{1}{2}}{|A^{-\hat{\alpha}}|^\frac{1}{2}}\exp\left(\frac{1}{2\beta^{2}}(v_0A^\alpha v_0-v_0A^{\hat{\alpha}} v_0)\right)$$
	
	\subsection{Sampling for Case study 2}
	
	We now look to the problem in section~2.3. The underlying dynamics are linear and the noise is Gaussian, so Kalman filtering provides efficient and quick estimates of $v_t$ given $(y_n)_{n\delta\leq t}$ and $\phi, \theta$. However estimating the system's parameters $(\phi, \theta)$ is a challenging problem. This study extends the work in \cite{sigrist_stochastic_2015} by also inferring $\alpha$ and hence redesigning the hyperpriors. 
	Compared to the previous case study, the use of a SPDE in the modelling means that after discretising in time one has a higher dimensional Gaussian prior on $\{\epsilon_t\}_{t\geq0}^T$. We will follow \cite{sigrist_spate_nodate} for sampling $p(v|y, \phi,\theta)$ using the forward filtering backward sampling (FFBS) algorithm \cite{fruhwirth-schnatter_data_1994}. Whilst this is not the pCN algorithm, it uses Kalman filtering recursions and achieves sufficiently fast mixing.
	
	In \cite{sigrist_spate_nodate} smoothness was deterministically fixed to $\alpha=2$ and was known. Also improper priors were used, which would cause unidentifiability issues for varying $\alpha$. Redesigning this requires the prior parameters for the remaining $\theta$ to be rescaled according to values of $\alpha$. So we will set these priors instead to: $\rho_{0}\sim\mathcal{U}[0.01, 5\cdot(\alpha-d/2)^{-2}],\;\sigma^{2}\sim\text{InvGamma}(1, 1),\;\gamma\sim\text{InvGamma}(5, 5),\;\zeta\sim\text{InvGamma}(1, 1),\;\tau^2\sim\mathcal{U}[0,100]$. We focused our efforts on the noise parameters since they are the most sensitive to changes in $\alpha$. The boundaries on the uniform prior of $\rho_0$ were chosen so that $\rho_0$ would be identifiable from its corresponding Matérn field. As for $\sigma^2$ we chose an Inverse Gamma hyperprior as it could be used in a conjugate relationship to sample directly from it. We also chose Inverse Gamma hyperpriors for $\gamma$ and $\zeta$ to allow the whole state space to be explored whilst penalising large moves. The rest are as in \cite{sigrist_stochastic_2015}: $\mu_x, \mu_y\sim\mathcal{U}[-0.5,0.5],\;\psi\sim\mathcal{U}[0,\pi/2],\;\rho_1\sim\mathcal{U}[0,100]$. We choose again a weakly informative prior on $\alpha$ to highlight the advantages of estimating it in a challenging case: $\mathcal{U}[1,4]$.
	
	\section{Numerical results\texorpdfstring{\protect\footnote{The code for the experiments can be found in \url{https://github.com/bs519/Hyperpar_est}.}}{0}}\label{sec_study}
	
	\subsection{Case study 1: inverse problem for Navier-Stokes}
	
	We are interested in comparing MwG where $\alpha$ is fixed and estimated for similar experiments to \cite{kantas_sequential_2014}, where $\alpha$ is assumed to be known and set to either $2$ or $2.2$. We produce figures in the study to compare the performance in initial condition $v_0$ and parameter $\beta^2$ estimation of the MwG assuming correct and incorrect values of $\alpha$ and estimating $\alpha$. For the PDE solver, we will use a spectral Galerkin method. We consider the finite dimensional projection onto the Hilbert space spanned by $\{\psi_{k}\}_{k\in\mathbb{L}_{n}}$ where $n\in \mathbb{N}\setminus\{0\}$ and $\mathbb{L}_{n} = \{(k_{1}, k_{2}) \in \mathbb{Z}^{2}\setminus\{0\}: \text{max}(|k_{1}|, |k_{2}|) < n/2\}$. We refer to the size of the implied mesh, $d_u$, as the dimensionality of u. We observe that $A$ is diagonalized in $U$ with eigenvalues $\{\lambda_{k}\}_{k\in \mathbb{Z}^{2}\setminus{\{0\}}}$, where $\lambda_{k} = |k|^{2}$. 
	Periodic boundary conditions allow the use of spectral Galerkin methods when solving the Navier-Stokes PDE numerically with Fast Fourier Transforms (FFTs). When facing different boundary conditions or forcing, we would use Finite Elements techniques or other methods \cite{girault_finite_1986}. Convolutions arising from products in the nonlinear term are computed via FFTs on our mesh with additional antialiasing using double-sized zero padding. In addition, exponential time differencing is used as in \cite{cox_exponential_2002} to deal with the resulting stiff system, whereby an analytical integration is used for the linear part, together with an explicit numerical Euler integration scheme for the nonlinear part.
	
	For the mesh $n={32}$, the number of observations is set to $T=5$ and for the observation grid $\Upsilon=16$. We have set $f(x)=\nabla^\bot \cos((5, 5)'\cdot x),\; \tau^2 =0.2$. We consider $N= 8\cdot10^5$ iterations and have found the optimal learning rate to be around $0.999$ and $0.998$ in a stationary ($\eta=0.1, \delta=1$) and "chaotic" regime ($\eta=0.02, \delta=0.02$) respectively. For the true initial condition and observations, we use a similar synthetic dataset, generated from a random sample of the prior with $\alpha=2.2,\, \beta^2=1.2$. We use $\pi(\beta^2)= \text{IG}(1.5,2.5)$.
	
	\subsubsection*{Stationary regime}
	
	\begin{figure}[h]
		\centering
		\includegraphics[width=\textwidth]{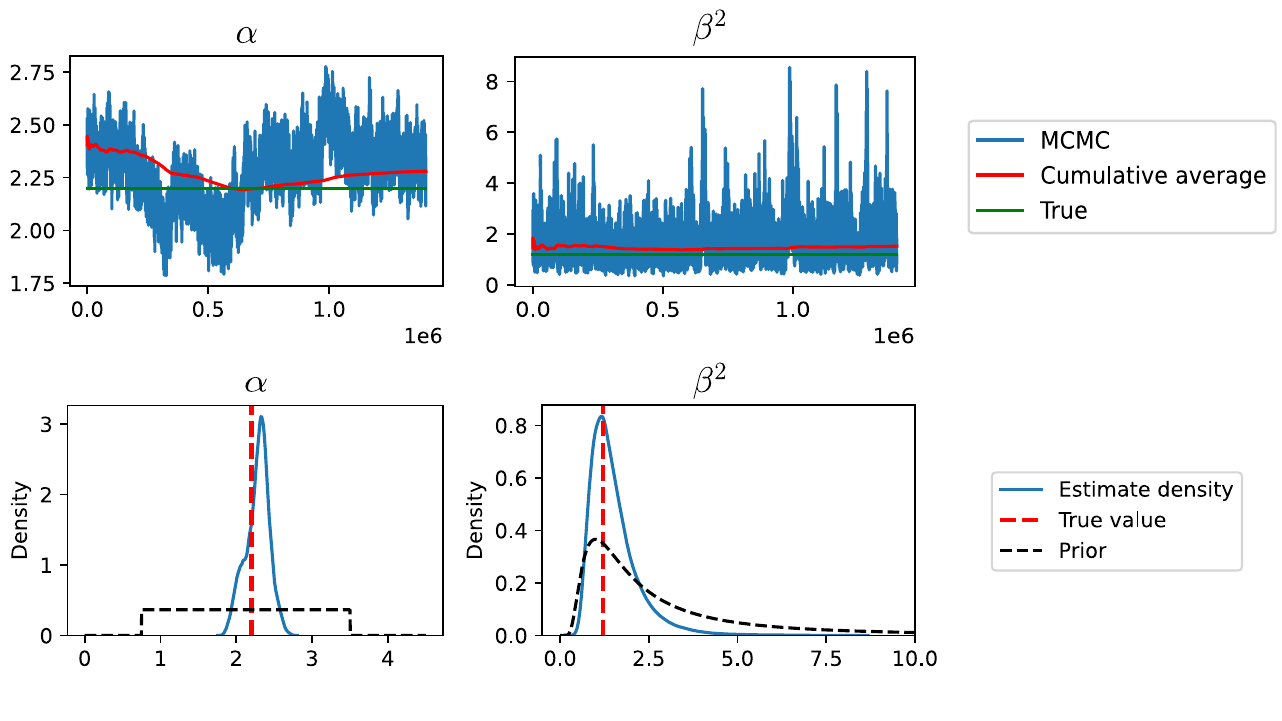}
		
		\caption{(Stationary regime) Trace (left) and density (right) plots of prior parameters. The trace plots show the MCMC estimates (blue) captures the true value (green) in its distribution, with its cumulative average (red) converging to a value close to the truth. Similarly the density plots display the posterior's estimate (blue) peaking around the true value (red) of both parameters. The prior corresponds to the black dashed line. For both parameters, the true value lies in the support of the densities of the estimates.
		}
		\label{fig:par_dens}
	\end{figure}
	
	\begin{figure}[h]
		\centering
		\includegraphics[width=\textwidth]{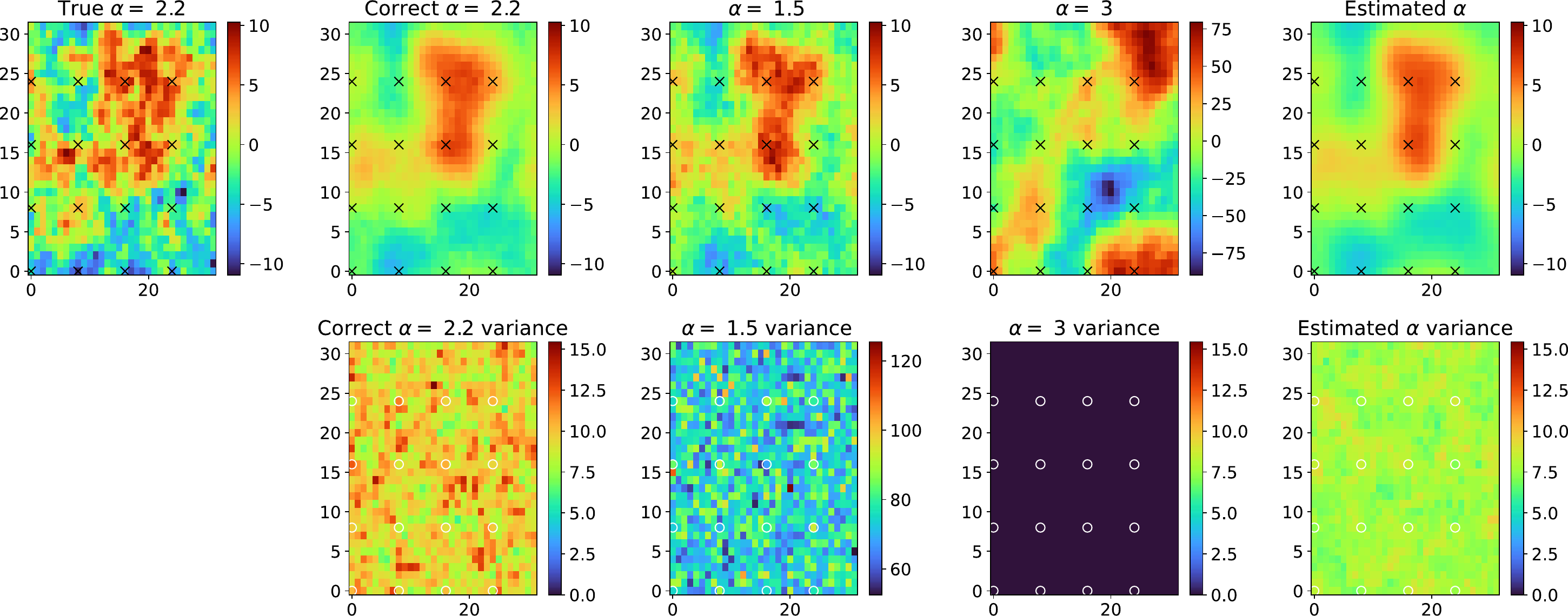}
		\caption{(Stationary regime) Heat map of the vorticity (top) of the initial condition $v_0$ with the truth (far left) and MCMC estimates: knowing the true $\alpha=2.2$ value (centre left), underestimating with $\alpha=1.5$ (centre), overestimating with $\alpha=3$ (centre right) and estimating $\alpha$ (far right). The crosses and circles indicate the positions where the vector field is observed. Heat map of the variance of each estimate (bottom). Note the scale of the variance being inversely proportional to $\alpha$. The algorithms knowing the true $\alpha$ and estimating it via MwG exhibit similar behaviour.}
		\label{fig:vortstat}
	\end{figure}
	
	The diagnostics plots for the different MwGs are in the Appendix in section \ref{diagnostics}. 
	PCN and MwG run times are shown in Table \ref{tableexp1}. We remark that as expected the runtimes are close to that of Algorithm \ref{Alg. MCMC}, with $\alpha, \beta^2$ known. 
	In Fig. \ref{fig:par_dens} we plot the trace plots and densities of our parameter estimation and the results prove the success of the parameter estimation even with uninformative priors. We now study the algorithm's uncertainty quantification of the initial condition. In Fig. \ref{fig:vortstat} we plot the vorticity of the initial condition's estimates against the true value. We observe similar estimates from the algorithms knowing the correct $\alpha$ and estimating it. The estimates recover the true initial condition relatively accurately, albeit giving a smoother estimate.

	We now focus on the variance of the estimates' vorticity to fully evaluate the algorithm's uncertainty quantification. We see from the scales that the variance is highly impacted by the choice of $\alpha$. We recover the order of variance when the true $\alpha$ is known only by estimating $\alpha$. When $\alpha$ is over and underestimated the uncertainty is under and over estimated, respectfully. Practitioners often choose to sacrifice uncertainty quantification for improved point estimates and vice-versa. This could be used within our framework by adjusting the priors of $\alpha$. Note however we will use a high viscosity and this lowers the system's energy, and in doing so, this stationary regime renders our variance estimation less crucial and the predictive power of all the algorithms similar after a few timesteps. We repeat this experiment below for the "chaotic" regime with lower viscosity and more frequent and informative observations to investigate the differences.
	
	\subsubsection*{Chaotic regime}
	
	\begin{figure}[htbp]
		\centering
		\includegraphics[width=\linewidth]{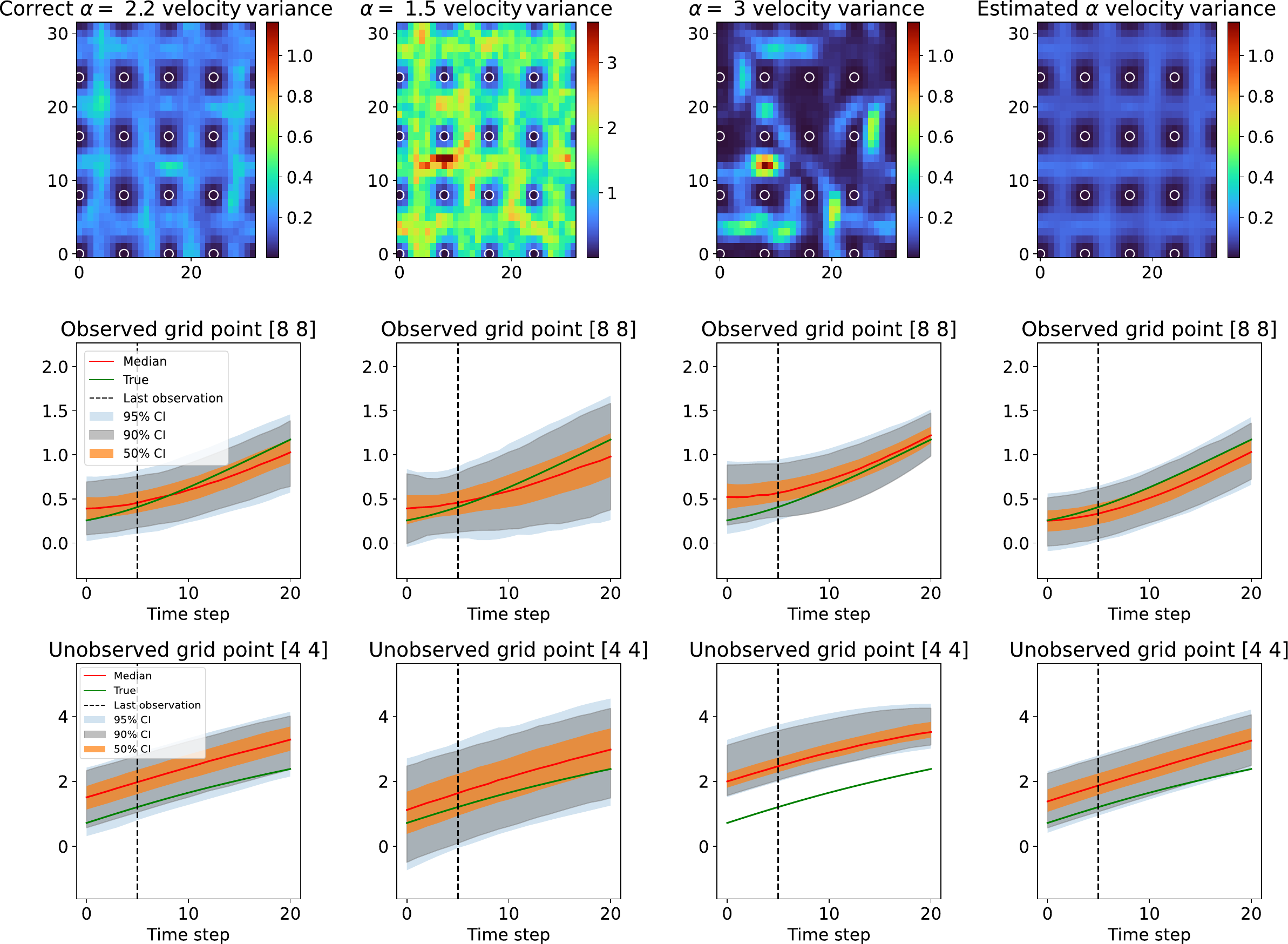}
		\caption{(Chaotic regime) Heat map of the variance of the velocity squared of $v_0$ vs $\alpha$ (top), as previously arranged, with: $\alpha=2.2$ value (far left), $\alpha=1.5$ (centre left), $\alpha=3$ (centre right) and estimating $\alpha$ (far right). The white circles in the top panels indicate the positions where the vector field is observed. We also plot the trajectory of the velocity squared of those estimates at observed (middle) and unobserved (bottom) locations. The vertical dashed line indicates the time of the last observation. Note the scale of the variance of the prediction estimates being inversely proportional to $\alpha$.
		}
		\label{fig:predpoststat_chaot}
	\end{figure}
	
	We obtain similarly good results in parameter estimation as in the stationary case, see Fig. \ref{fig:par_dens_chaot}. This is also the case for the vorticity point estimates shown in Fig. \ref{fig:vortstat_chaot}. Fig. \ref{fig:beta_chaotic} illustrates that when $\alpha$ is misspecified, the estimate of $\beta^2$ will compensate in order to recover the correct variance, leading to poor parameter estimation. These plots can be found in the Appendix in section \ref{chaotic_plots}. In Fig. \ref{fig:predpoststat_chaot} we present the results similarly to Fig. \ref{fig:vortstat} previously. The scale of the variance of the vorticity estimates is as before at unobserved locations, however there is spatial correlation introduced by the observations. 
	The likelihood is peaky around the informative observation locations, which reduces the variance of the posterior of the velocity at these locations, as well as the posterior's dependence on the prior. For higher values of $\alpha$ this spatial correlation is more pronounced as the corresponding prior variance is smoothed. So a higher $\alpha$ leads to a less informative prior variance, additionally increasing the posterior's dependence on the likelihood. This is why the variance is impacted by $\alpha$ as it was in the stationary setting only at unobserved locations. This can be additionally seen in the plots of the trajectories of the velocity's estimates at observed times. Due to the chaotic regime having high energy, the trajectory plots show that the difference in variances does not vanish in time and in turn impacts the estimate's predictive power. $\alpha=3$ leads to an underestimated variance at unobserved locations: the confidence interval does not capture the true value after very few time steps past the observations. At observed locations the forecast performs well, meaning the likelihood is enough to correct the misspecified prior. As for $\alpha=1.5$, the variance is overestimated leading to poor uncertainty quantification at all times. At observed locations its forecast performs well until the last observation, its predictive power degrades quickly as the variance explodes. Then MwG with $\alpha=1.5$ performs poorly at uncertainty quantification, particularly for unobserved states and prediction. This shows that even where the likelihood is most influential, the prior still influences the algorithm's predictive power. Finally the MwG estimating $\alpha$ reproduces similar performance to the algorithm assuming its correct value and performs similarly well even in prediction.
	
	\subsection{Case study 2: advection-diffusion SPDE}
	
	\begin{figure}
		
		\includegraphics[width=\textwidth]{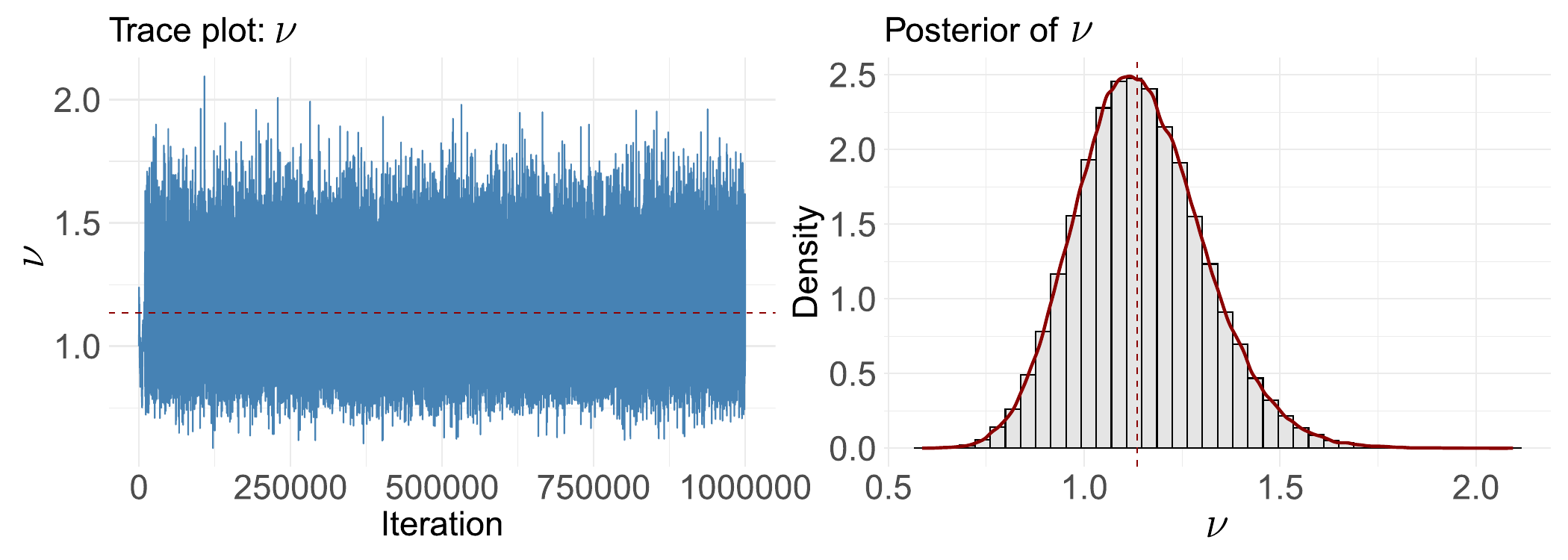}
		
		\caption{Trace (left) and density (right) plot of smoothness parameter estimates in case study 2. The dashed line corresponds to the median estimate $\approx1.1$, close the true value generating the data $\alpha=\nu+1=2$.}
		\label{fig:par_dens_R}
	\end{figure}
	
	\begin{figure}[htbp]
		\begin{tabular}{cc}
			
			\begin{subfigure}{0.5\textwidth}
				\includegraphics[width=\textwidth]{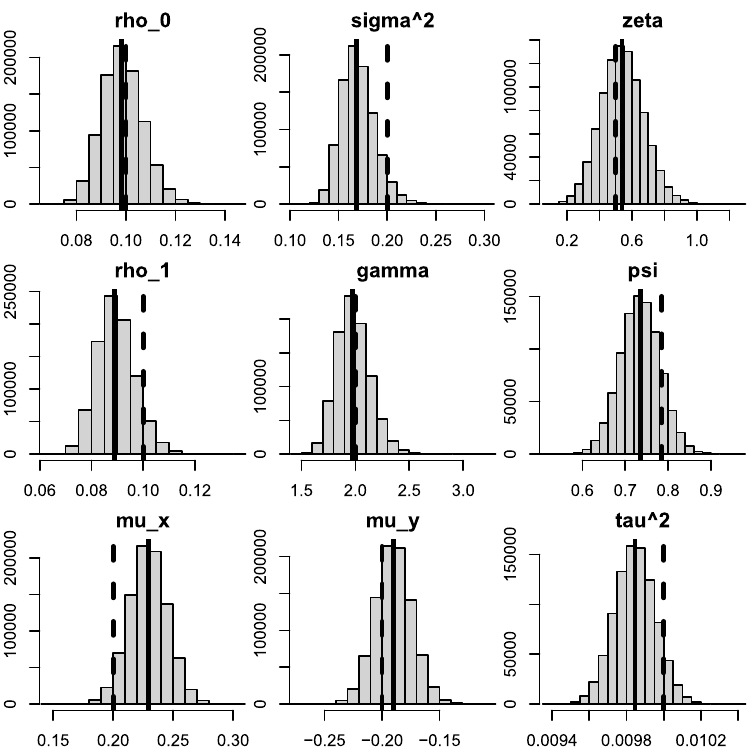}
				\caption{$\alpha=2$}
			\end{subfigure}
			&
			\begin{subfigure}{0.5\textwidth}
				\includegraphics[width=\textwidth, height=7.7cm]{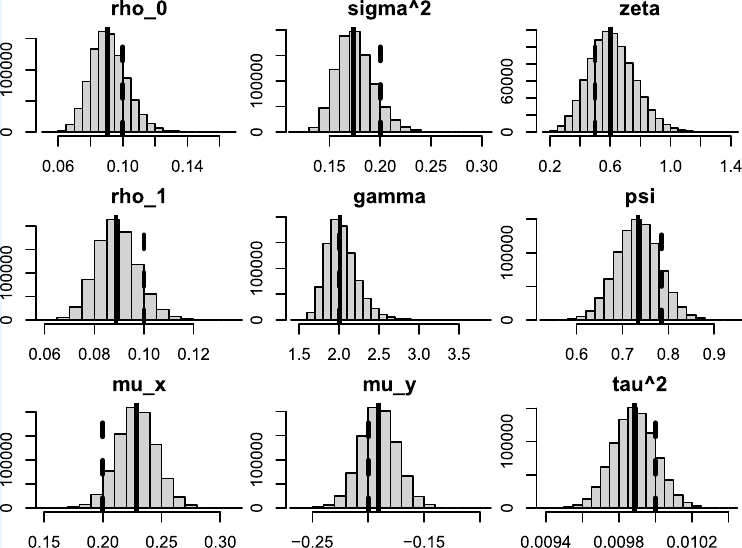}
				\caption{Estimated $\alpha$}
			\end{subfigure}
			\\
			\begin{subfigure}{0.5\textwidth}
				\includegraphics[width=\textwidth]{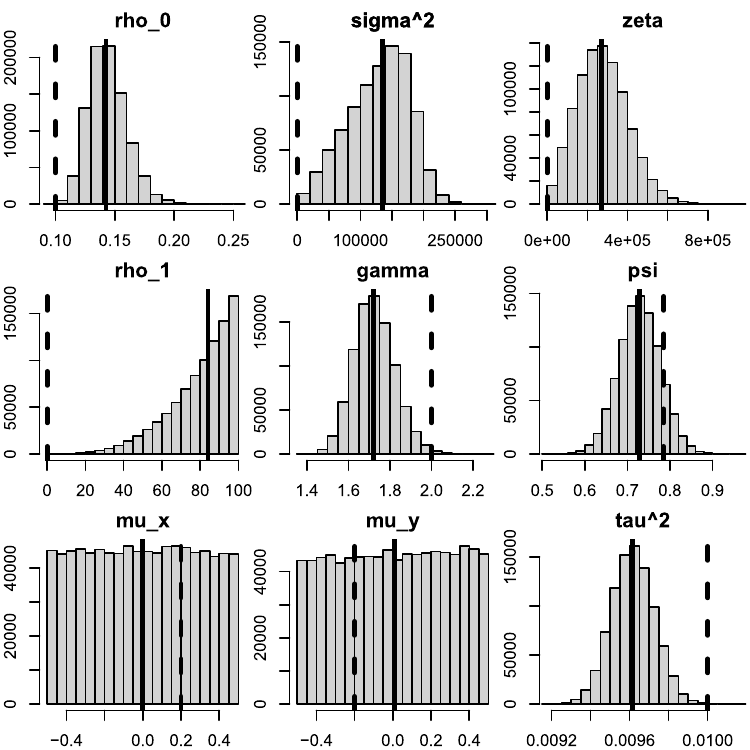}
				\caption{$\alpha=1.5$}
			\end{subfigure}
			
			&
			\begin{subfigure}{0.5\textwidth}
				\includegraphics[width=\textwidth]{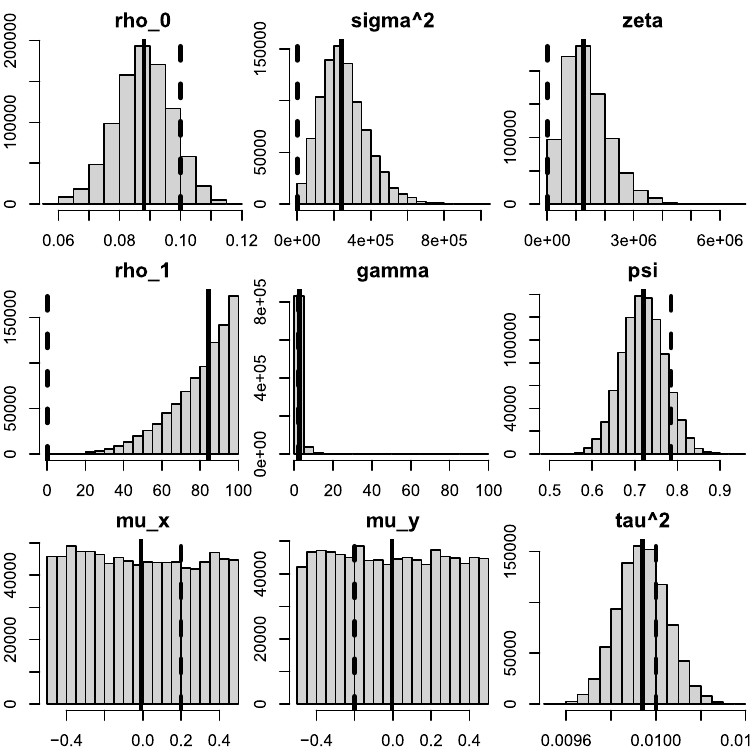}
				\caption{$\alpha=2.5$}
			\end{subfigure}
		\end{tabular}
		
		\caption{Histogram of the parameters estimated when the algorithm knows the true $\alpha = 2$ value (top left), underestimates $\alpha = 1.5$ (bottom left), overestimates $\alpha = 2.5$ (bottom right) and estimates $\alpha$ (top right). The dashed line indicates the true value and the full line represents the median of the samples. Assuming the wrong value of $\alpha$ leads to very poor parameter estimation with parameter samples for $\sigma^2, \zeta, \rho_1, \gamma, \mu_x, \mu_y, \tau$ in non-representative values. Samples for $\mu_x, \mu_y$ are still distributed according to their prior meaning no information was gained from the data for $\mu$. However estimating $\alpha$ and knowing the correct $\alpha$ allows for similar accurate parameter estimation results, as the distributions of their estimates are centered around or capture well the true value.}
		\label{fig:par_hist}
	\end{figure}
	
	We reuse the code and example provided in \cite{sigrist_spate_nodate} using the R package 'spate'. The discretisation grid is of size $n^2$ with $n=32$. We set $T=20$ observations, generated from the whole grid ($\Upsilon=n^2$) with Gaussian noise $\tau^2=0.01$. 
	$\xi$ is generated from parameters $\rho_0=0.1,\sigma^2=0.2,\zeta=0.5,\rho_1=0.1, \gamma=2,\alpha=\pi/4,\mu_1=0.2,\mu_2=-0.2$. We pick $N=10^6$ with $10\%$ burn-in and $5000$ to be both the burn-in period and the minimal number of MCMC samples after this burn-in to estimate the proposal matrix. We pick as starting values $\rho_0=0.2,\sigma^2=0.1, \zeta=0.25,\rho_1=0.2,\gamma=1,\alpha=0.3,\mu_1=0,\mu_2=0,\tau^2=0.005$ and initiate $\alpha$ correctly. As before we compare the performance of the algorithm additionally estimating the smoothness to the standard algorithm assuming correct and incorrect values of $\alpha$. Results on $\alpha$ and the remaining parameters are shown in Fig. \ref{fig:par_dens_R} and Fig. \ref{fig:par_hist} respectively. The estimation of $\alpha$ as $2.1$ is close to the true value $2$ and the parameter space is well explored. Assuming the wrong value of $\alpha$ leads to very poor parameter estimation with parameter samples for $\sigma^2, \zeta, \rho_1, \gamma, \mu_x, \mu_y, \tau$ in non-representative values. However estimating $\alpha$ allows for parameter estimation results very close to the accurate performance of the standard algorithm knowing the correct $\alpha$, as the distributions of their estimates are centered or capture well the true value. This is additionally supported by the parameters' trace plots in Fig. \ref{fig:par_trace} in section~\ref{SAD_plots} in the Appendix.
	
	\section{Conclusion}\label{sec_conclusion}
	
	In this work, we consider two case studies that share common characteristics with operational data assimilation problems. In both cases, we have illustrated the merits of estimating the smoothness parameter together with other parameters of the prior and likelihood as well as noticed performance similar to knowing the true value of $\alpha$. At the same time, we showed how misspecifying $\alpha$ can lead to poor uncertainty quantification or estimation of other parameters in $\phi, \theta$. Our case studies covered different Gaussian priors (Matern/Laplacian), different dynamics (linear, nonlinear stationary or chaotic), observation regimes (dense or sparse observation grids). So we believe using hierarchical modelling for $\phi, \theta$ together with MwG can be very helpful in practice. Even when little is known about the smoothness of the state, it can be inferred and by exploring the parameter space it adds flexibility to the model. 
	We have also established that in certain cases practitioners may choose to sacrifice uncertainty quantification for improved point estimates and vice-versa. In real NWP operational problems the state dimension may reach $\mathcal{O}(10^7-10^9)$, whereas our proof of concept looks at $\mathcal{O}(10^3)$. Still we highlight that the overhead of estimating $\alpha$ as well as the rest of $\theta$ and $\phi$ can be minimal. 
	Finally, this approach could be generalized/extended to more elaborate priors: e.g. Matern with Cauchy noise \cite{suuronen_cauchy_2022} or the more general $\alpha$-stable distributions \cite{suuronen_bayesian_2023}, with Student's $t$ priors as in \cite{senchukova_bayesian_2024} or sparsity-promoting priors as in \cite{emzir_non-stationary_2020,monterrubio-gomez_posterior_2020}.
	
	\section*{Acknowledgements}
	B.S. was supported by the Additional Funding Programme for Mathematical Sciences, delivered by EPSRC (EP/V521917/1) and the Heilbronn Institute for Mathematical Research.
	
	\section*{Conflict of interest}
	
	The authors state that none of the work described in this publication appears to have been influenced by any known conflicting financial motives or personal ties.
	
	\newpage
	
	\bibliographystyle{siam}
	\bibliography{LSR}
	
	\newpage
	
	\appendix
	
	\section{The Navier Stokes equation}\label{NS appendix}
	
	We introduce here the Navier-Stokes equation for incompressible flow. Let $v(x, t)$ denote the velocity vector field at point $x$ and time $t$. The domain we consider is a 2D Torus $\mathbb{T}=[0,2\pi]^{2}$ with initial flow $v_0\in \mathbb{T}$ and let $v : \mathbb{T} \times [0, \infty) \rightarrow \mathbb{R}^{2}, v(x, t)=(v_{1}(x, t),v_{2}(x, t))^T$. The dynamics for $v$ are as follows:
	
	\begin{equation}\label{NS eq}
		\partial_{t}v - \eta\Delta v +(v \cdot\nabla) v = f -\nabla p , \qquad v(x, 0) = v_0(x)
	\end{equation}
	\begin{equation}
		\nabla\cdot v =0, \int_\mathbb{T} v_{j}(x, \cdot)dx =0,\; j =1, 2,
	\end{equation}
	where $\eta>0$ is the viscosity parameter, $p : \mathbb{T} \times [0, \infty) \rightarrow \mathbb{R}$ is the pressure function, and $f : \mathbb{T} \rightarrow \mathbb{R}^{2}$ is an exogenous time-homogeneous forcing. 
	For simplicity we consider the mean flow to be 0 and assume periodic boundary conditions: $v_j(\cdot, 0,t)=v_{j}(\cdot, 2\pi, t),\,  v_{j}(0, \cdot,t)=v_{j}(2\pi, \cdot,t),\, j =1, 2\,\forall t\geq0$. Here $v_0$ is the initial velocity field to be estimated. We now define its domain as the following Hilbert space $\mathbb{V}$ with closure $V$ with respect to $L^{2}(\mathbb{T})^{2}$ norm:
	$$\mathbb{V}:= \left\{2\pi\text{-periodic trigonometric polynomials }u: \mathbb{T}\rightarrow\mathbb{R}^{2}|\;\nabla\cdot u=0, \int_\mathbb{T} u(x)dx=0 \right\}.$$
	
	Let $P:(L^{2}(\mathbb{T}))^{2} \rightarrow V$ denote the Leray–Helmholtz orthogonal projector. $P$ projects $f$ onto the space of incompressible functions. In our context this means the pressure term in \eqref{NS eq} is removed. We obtain the equation as in \eqref{NS_ODE}, repeated below for convenience:
	\begin{equation}
		\frac{dv}{dt} + \eta Av + B(v, v)=P (f ) ,\;v(0) = v_0,
	\end{equation}	
	\noindent where $B(v,\tilde{v})= \frac{1}{2} P ((v \cdot \nabla) \tilde{v})+\frac{1}{2} P (( \tilde{v} \cdot \nabla) v)$ and $A = -P \Delta$ is the Stokes operator. If this set of equations is projected on a finite dimensional  subspace, then it is an ODE.
	
	\section{Conjugate prior derivation}\label{conj_prior}
	
	In this section we explicitly show the conjugate relationship of the parameter $\beta^2$, in Case study 1 used in section \ref{par. est. case 1}, exploiting the conjugate prior with Inverse Gamma distribution for the Gaussian likelihood. A similar conjugate relationship holds for $\sigma^2$ in case 2.
	
	\begin{align*}
		\pi (\beta^{2}|y, \alpha, v_0) & \propto \quad \frac{\sqrt{2}}{\sqrt{2\pi}^{d}|\beta|^{d}|A^{-\alpha}|^\frac{1}{2}}\exp(-\frac{1}{2\beta^{2}}v_0A^\alpha v_0) \cdot \frac{b^{a}}{\Gamma(a)}(\beta^{2})^{-a -1} \exp \left(-\frac {b }{\beta^{2}}\right)\\
		& \propto \quad \frac{b^{a}}{\Gamma(a)}(\beta^{2})^{-a -1}(\beta^{2})^{-\frac{d}{2}}\exp \left(-\frac {b}{\beta^{2}} -\frac{1}{2\beta^{2}}v_0A^\alpha v_0\right)\\
		& \propto \quad \frac{b^{a}}{\Gamma(a)}(\beta^{2})^{-(a +\frac{d}{2})-1}\exp \left(-\frac {b + \frac{1}{2}v_0A^\alpha v_0}{\beta^{2}}\right)\\
		& \propto \quad \frac{(b+ \frac{1}{2}v_0A^\alpha v_0)^{a+\frac{d}{2}}}{\Gamma(a+\frac{d}{2})}(\beta^{2})^{-(a +\frac{d}{2})-1}\exp \left(-\frac {b + \frac{1}{2}v_0A^\alpha v_0}{\beta^{2}}\right)
	\end{align*}
	
	\section{Case study 1: diagnostics and run times}
	
	\subsection{Diagnostics}\label{diagnostics}
	
	\begin{figure}[H]
		
		\begin{subfigure}{0.3\textwidth}
			\includegraphics[width=1.2\textwidth, height=4.2cm]{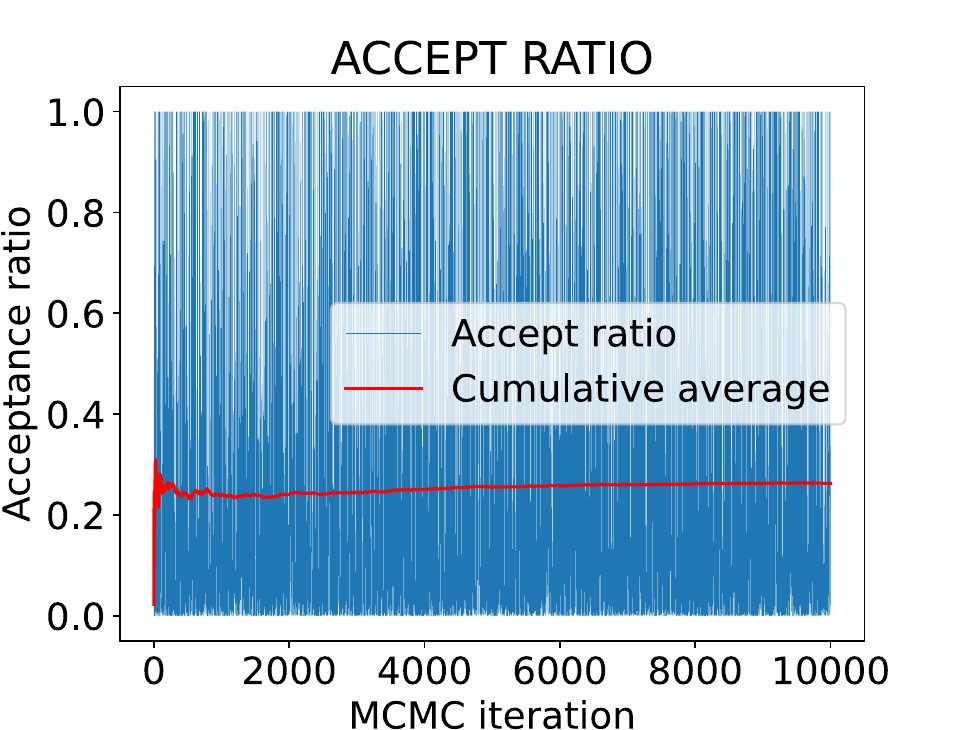}
			\caption{Acceptance ratio}
			\label{fig:accept_ratio}
		\end{subfigure}
		\hfill
		\begin{subfigure}{0.3\textwidth}
			\includegraphics[width=\textwidth, height=4cm]{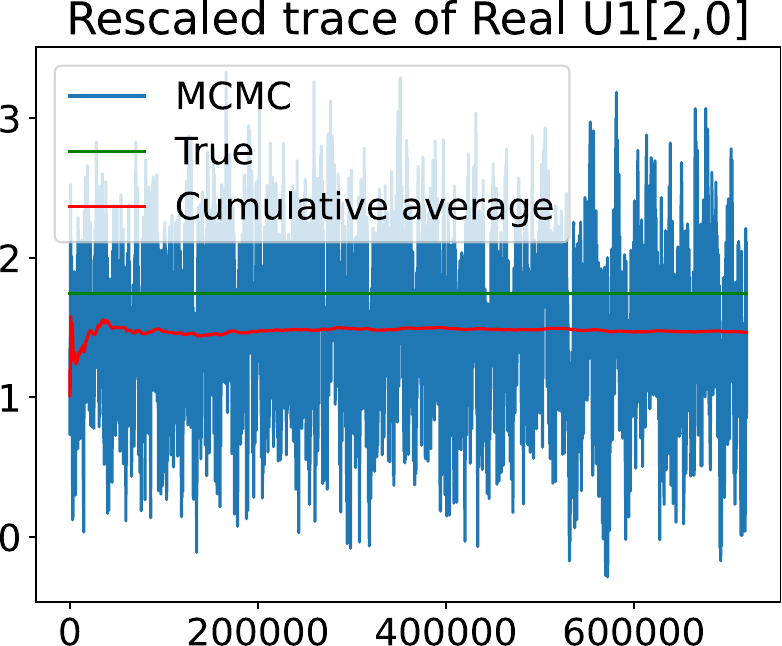}
			\caption{Velocity trace plot}
			\label{fig:trace_u1[2,0]}
		\end{subfigure}
		\hfill
		\begin{subfigure}{0.3\textwidth}
			\includegraphics[width=\textwidth, height=4cm]{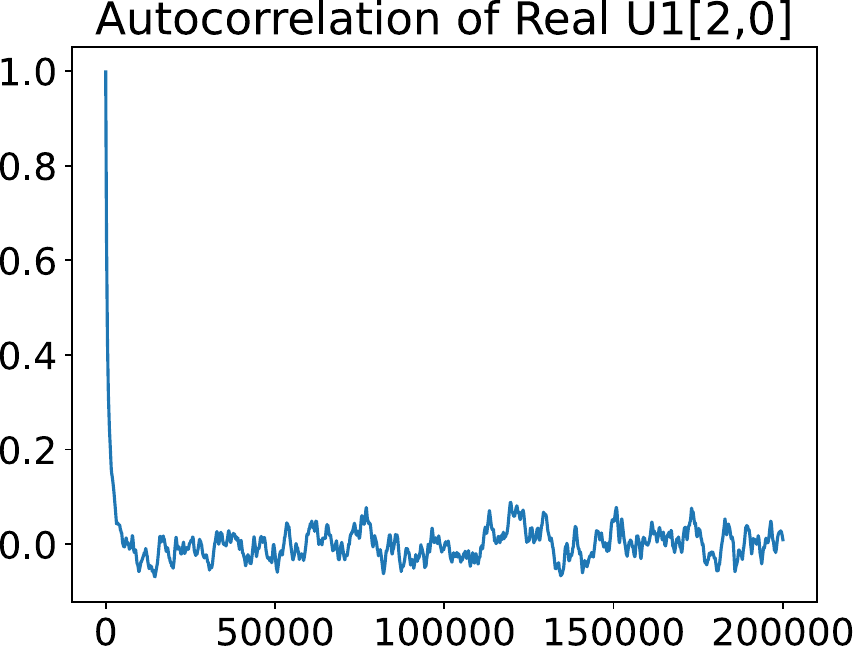}
			\caption{Autocorrelation plot}
			\label{fig:autocorr}
		\end{subfigure}
		
		\caption{pCN diagnostics: the cumulative average (red) of the acceptance ratio (left) is around $0.23$ as desired. The acceptance ratio (blue) does not alternate between 0 and 1 despite the blue background, it is diverse as desired. The trace plot (middle) of the real part of the velocity in the first dimension at the point [2,0] of our grid shows the MCMC estimate (blue) oscillates around the true value (green), with its cumulative average (red) converging to a value close to the truth. The autocorrelation plot of this quantity shows that the MCMC chain decorrelates after $1.5\cdot10^5$ iterations.}
		\label{fig:mcmc_plots}
	\end{figure}
	
	\begin{figure}[htbp]
		\includegraphics[width=\textwidth, height=9cm]{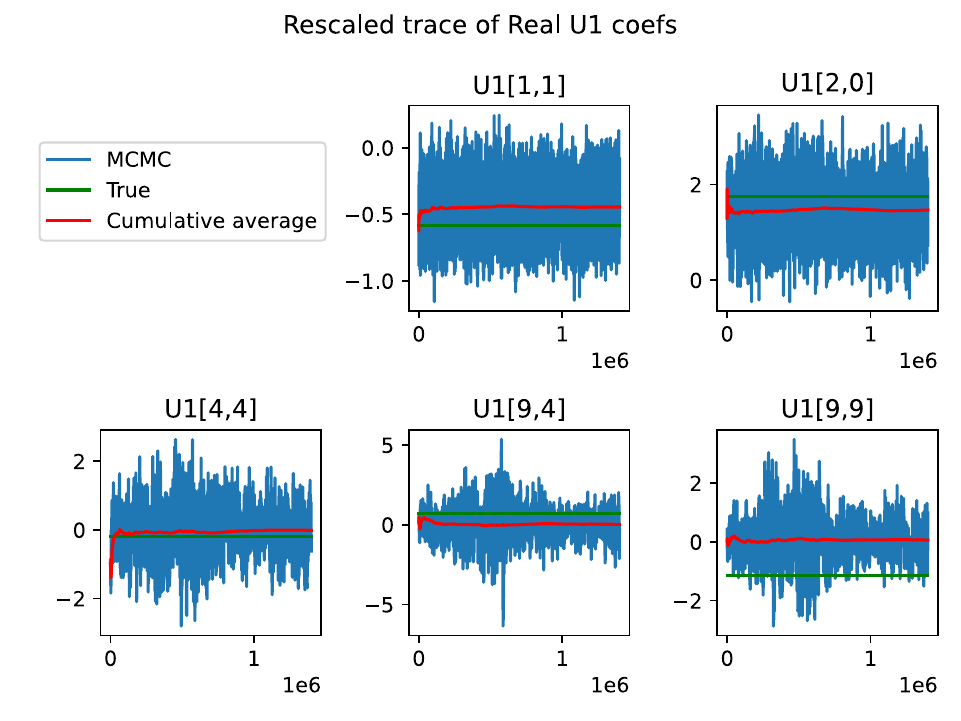}
		
		\caption{MwG estimating $\alpha$ diagnostics: the cumulative average of the acceptance ratio of the initial condition and of $\alpha$ is around $0.23$ as desired. The acceptance ratio is diverse as desired. The trace plot of the real part of the velocity in the first dimension at various points shows each MCMC estimate (blue) captures the true value (green) in its distribution, with its cumulative average (red) converging to a value mostly close to the truth.}
		\label{fig:trace_mwg}
	\end{figure}

	Fig. \ref{fig:mcmc_plots} diagnostics for pCN show that $N=8\cdot 10^5$ iterations are sufficient for the MCMC to provide good estimates. In Fig. \ref{fig:accept_ratio} we can observe that the acceptance ratio is stable after the burn-in period of $8\cdot10^4$ and between the desired values of 0.2 and 0.3. Focusing on the real value of a single Fourier coefficient, in Fig. \ref{fig:trace_u1[2,0]} the MCMC estimate performs well. Additionally its autocorrelation plot in Fig. \ref{fig:autocorr} suggests that despite high decorrelation lags, the high number of MCMC iterations will lead to a reasonable effective sample size. For MwG, we increase the number of MCMC moves to the traditional MCMC algorithm to take into account the Gibbs sampler's moves on the hyperparameters: $N=1.6\cdot10^6$ for $p_{v_0}= p_{\beta^2}=\frac{1}{3}$ and $N=2.4\cdot10^6$ for $p_{v_0}= p_{\beta^2}= p_\alpha=\frac{1}{3}$. To fully observe the behaviour of the parameter estimation, we use longer runs: $N=6\cdot 10^6$. This is solely for our study and would not be needed in practice. We found the optimal step-size for $\alpha$ to be $0.96$ for both regimes. Trace plots of the velocity estimates are shown in Fig. \ref{fig:trace_mwg}. 
	
	\subsection{Chaotic regime plots}\label{chaotic_plots}
	
	\begin{figure}[H]
		\centering
		\includegraphics[width=\textwidth]{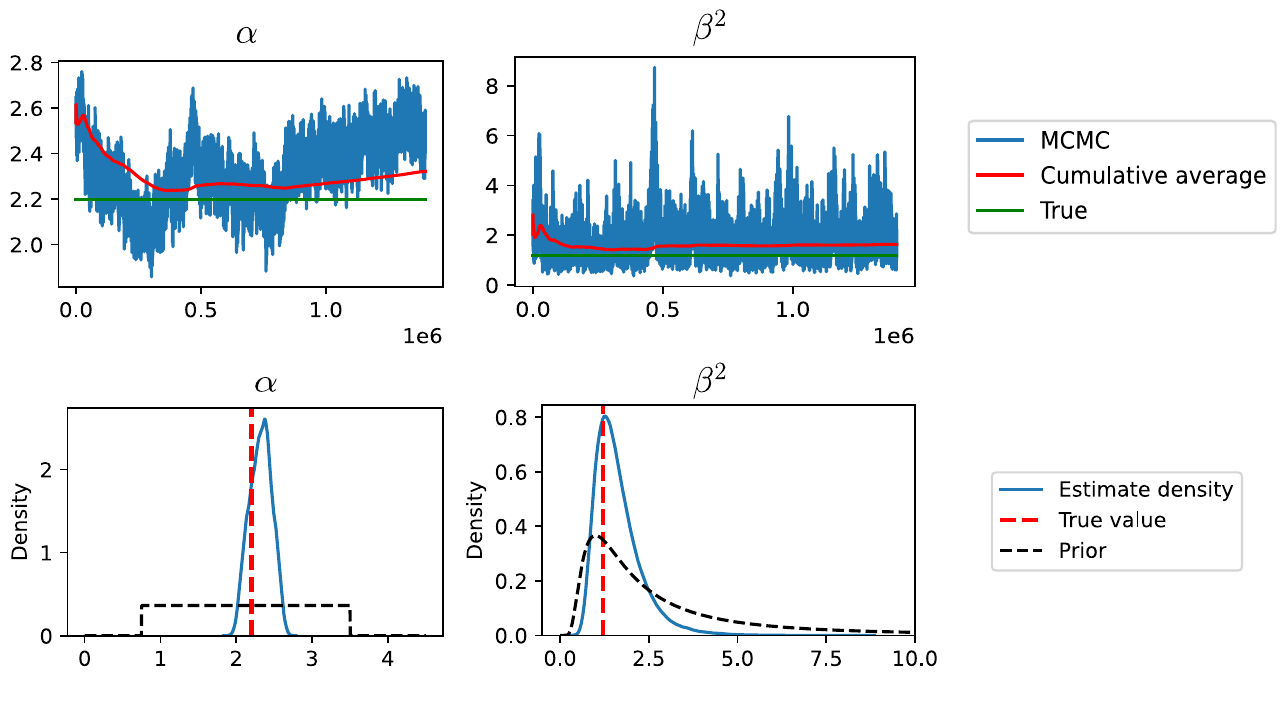}
		
		\caption{Trace (left) and density (right) plots of prior parameters. The trace plots show the MCMC estimates (blue) captures the true value (green) in its distribution, with its cumulative average (red) converging to a value close to the truth. Similarly the density plots display the posterior's estimate (blue) peaking around the true value (red) of both parameters. The prior corresponds to the black dashed line. For both parameters, the true value lies in the support of the densities of the estimates. For $\beta$, the challenging heavy-tailed Gamma prior is shrunk around the true value to form the posterior.}
		\label{fig:par_dens_chaot}
	\end{figure}
	
	\begin{figure}[h]
		
		\includegraphics[width=\textwidth]{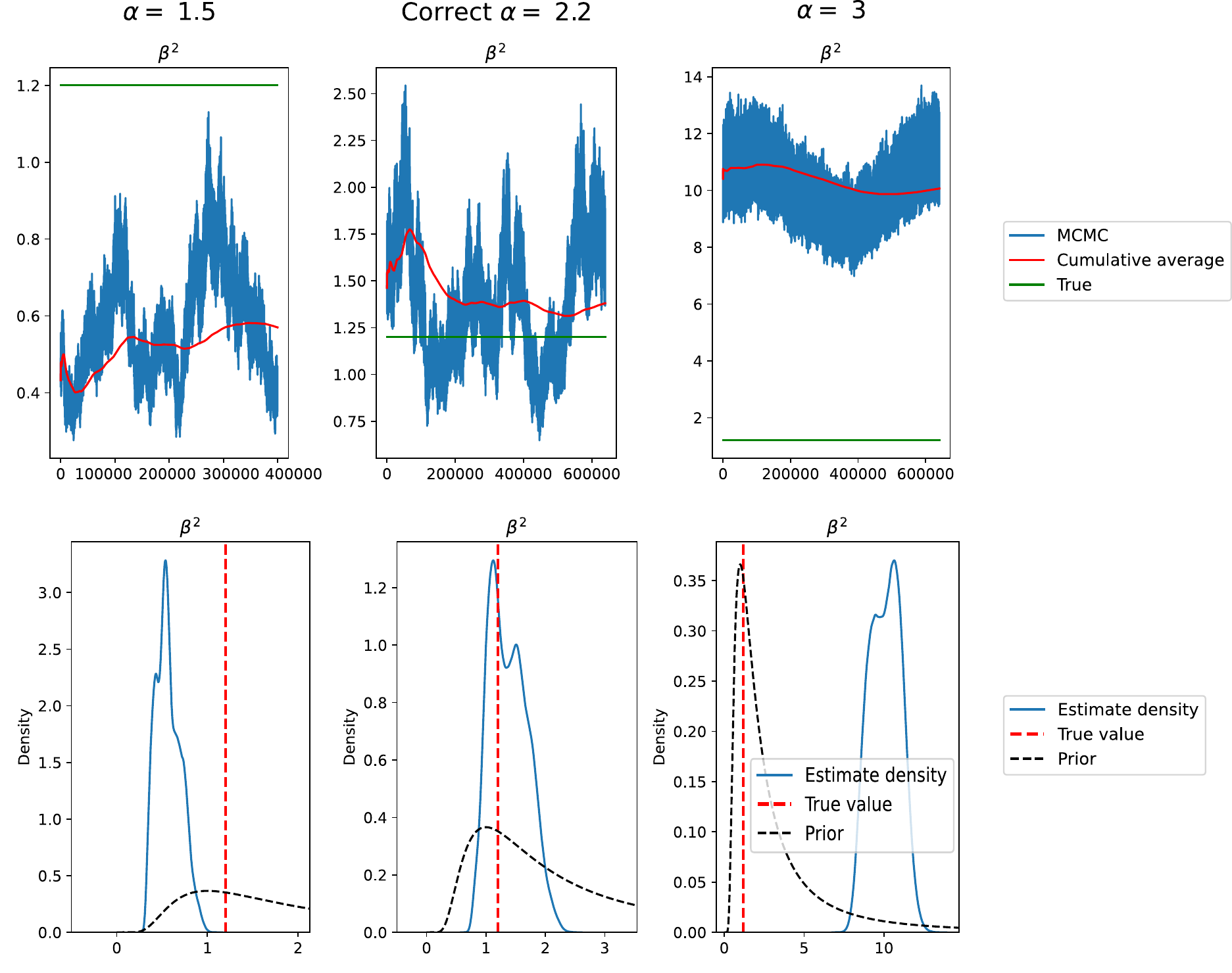}
		\caption{$\beta^2$ estimates: Trace (top) and density (bottom) plots of prior parameter $\beta^2$ for MwG: knowing the true $\alpha = 2.2$ value (centre), underestimating with $\alpha = 1.5$ (left), overestimating with $\alpha = 3$ (right). The trace plots show the estimates (blue), with its cumulative average (red) and the true value (green). The density plots display the posterior’s estimate (blue) peaking around the true value (red) of both parameters. The prior corresponds to the black dashed line. The true value lies in the support of the density of the estimates for the correct $\alpha$. However this density when $\alpha=1.5$ underestimates $\beta^2$ and conversely for $\alpha=3$.}
		\label{fig:beta_chaotic}
	\end{figure}
	
	\begin{figure}[H]
		\centering
		\includegraphics[width=\textwidth]{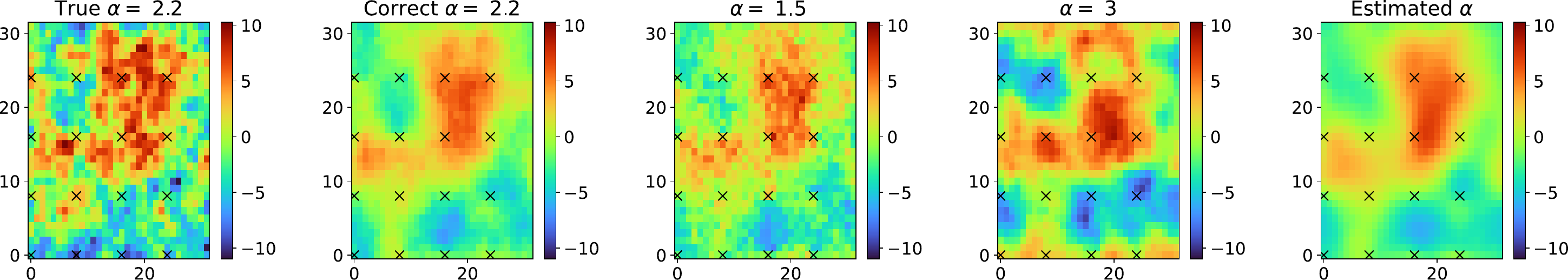}
		\caption{Vorticity of the initial condition $v_0$ with the truth (far left) and MCMC estimates: knowing the true $\alpha=2.2$ value (centre left), underestimating $\alpha=1.5$ (centre), overestimating $\alpha=3$ (centre right) and estimating $\alpha$ (far right). The crosses indicate the positions where the vector field is observed. We observe similar estimates from the algorithms knowing the correct smoothness and estimating it, albeit giving a smooth estimate.}
		\label{fig:vortstat_chaot}
	\end{figure}

	\subsection*{Run times}\label{tableexp1}
	
	\begin{table}[H]
		\begin{tabular}{||c| c c c c c ||}
			\hline
			Algorithm & Regime & Model & Configuration & Execution time & PDE count \\ [0.5ex] 
			\hline\hline
			pCN & Stationary & N-S & $N=8\cdot10^5$ & 25 hours & $8\cdot10^7$\\
			\hline
			MwG$^1$ & Stationary & N-S & $N=1.6\cdot10^6$ & 27.5 hours & $2\cdot10^8$\\
			\hline
			MwG$^2$ & Stationary & N-S & $N=6\cdot10^6$ & 85 hours & $2\cdot10^8$\\
			\hline
			pCN & Chaotic & N-S & $N=8\cdot10^5$ & 12.5 hours & $4\cdot10^6$\\ 
			\hline
			MwG$^1$ & Chaotic & N-S & $N=1.6\cdot10^6$ & 15 hours & $4\cdot10^6$\\
			\hline
			MwG$^2$ & Chaotic & N-S & $N=6\cdot10^6$ & 49 hours & $5\cdot10^6$\\
			\hline
			MwG & N/A & Stoch. AD & $\alpha:=2$ &  11 hours & N/A\\ 
			\hline
			MwG & N/A & Stoch. AD & $\mathcal{U}(1,4)$ & 12.5 hours & N/A\\
			\hline
		\end{tabular}
		\caption{Run times of the MwG algorithm estimating $\theta,\phi$ and of the pCN algorithm knowing $\theta, \phi$. In Case study 1, the MwG with $p_{v_0}=p_{\beta^2}=1/2$ and with $p_{v_0}= p_{\beta^2}= p_\alpha=1/3$, denoted respectively as MwG$^1$ and MwG$^2$, are slightly slower than pCN due to additional memory requirements for storing hyperparameter samples. Taking this into account and looking at the number of times the PDE solver was ran, the run times highlight that estimating $\theta$ does not significantly increase the run times over the known $\theta$ case. The computational run times are dominated by the runs of the PDE solver when evaluating the likelihood, which was confirmed when profiling the code.}
	\end{table}
	
	\newpage
	
	\section{Case Study 2: trace plots}\label{SAD_plots}
	
	\begin{figure}[h]
		\begin{tabular}{cc}
			
			\begin{subfigure}{0.5\textwidth}
				\includegraphics[width=\textwidth]{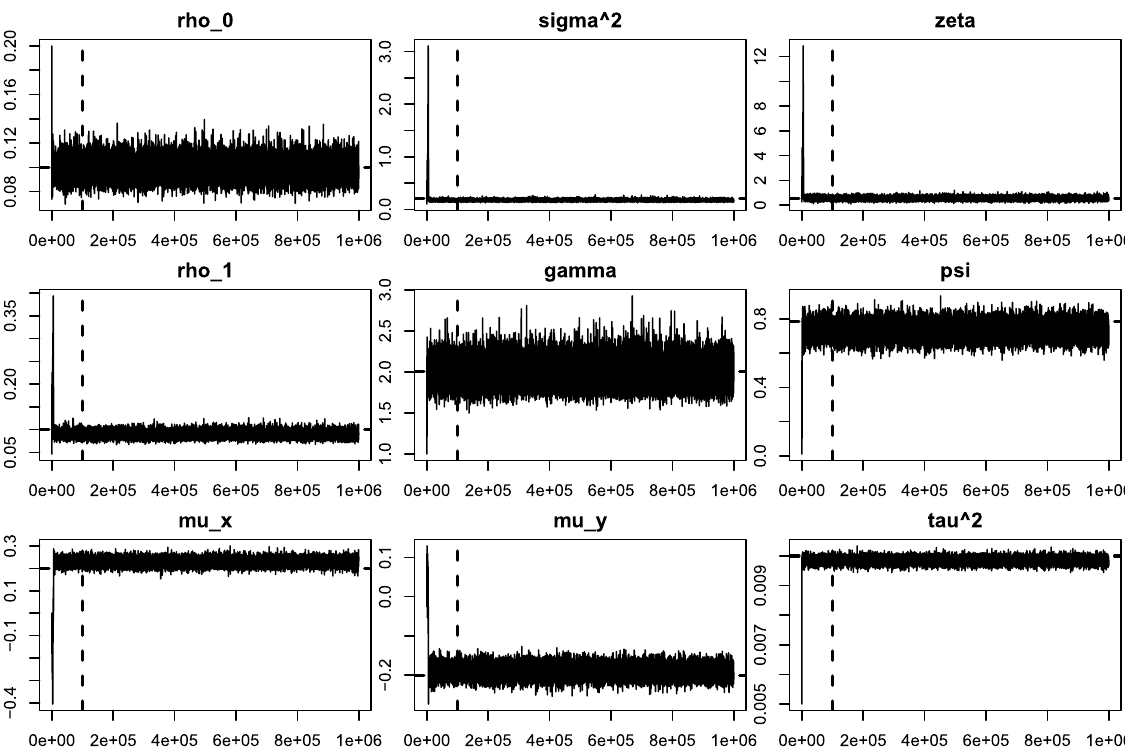}
				\caption{$\alpha=2$}
			\end{subfigure}
			&
			\begin{subfigure}{0.5\textwidth}
				\includegraphics[width=\textwidth, height=5.13cm]{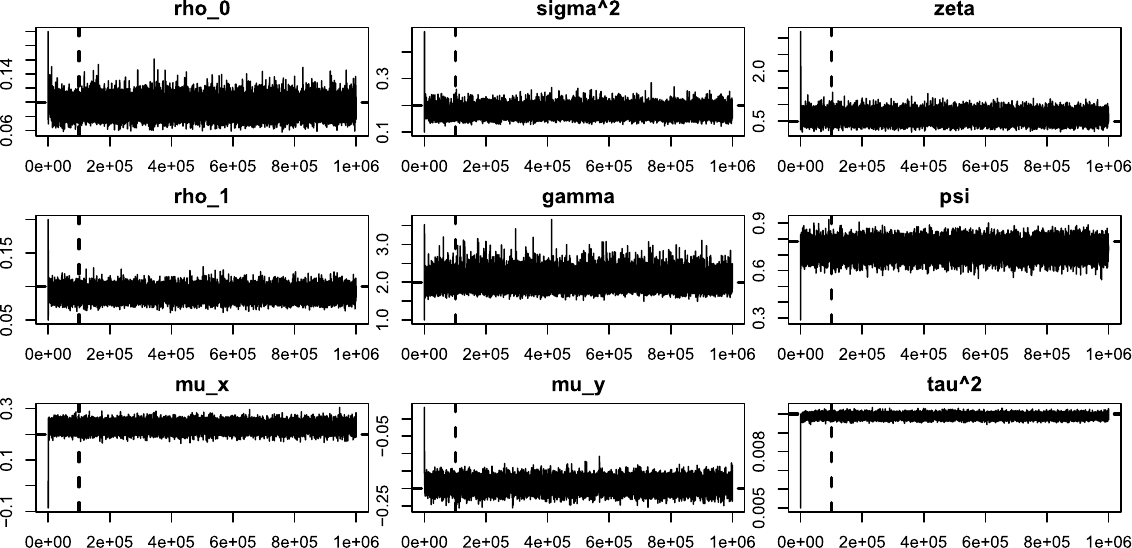}
				\caption{Estimated $\alpha$}
			\end{subfigure}
			\\
			\begin{subfigure}{0.5\textwidth}
				\includegraphics[width=\textwidth]{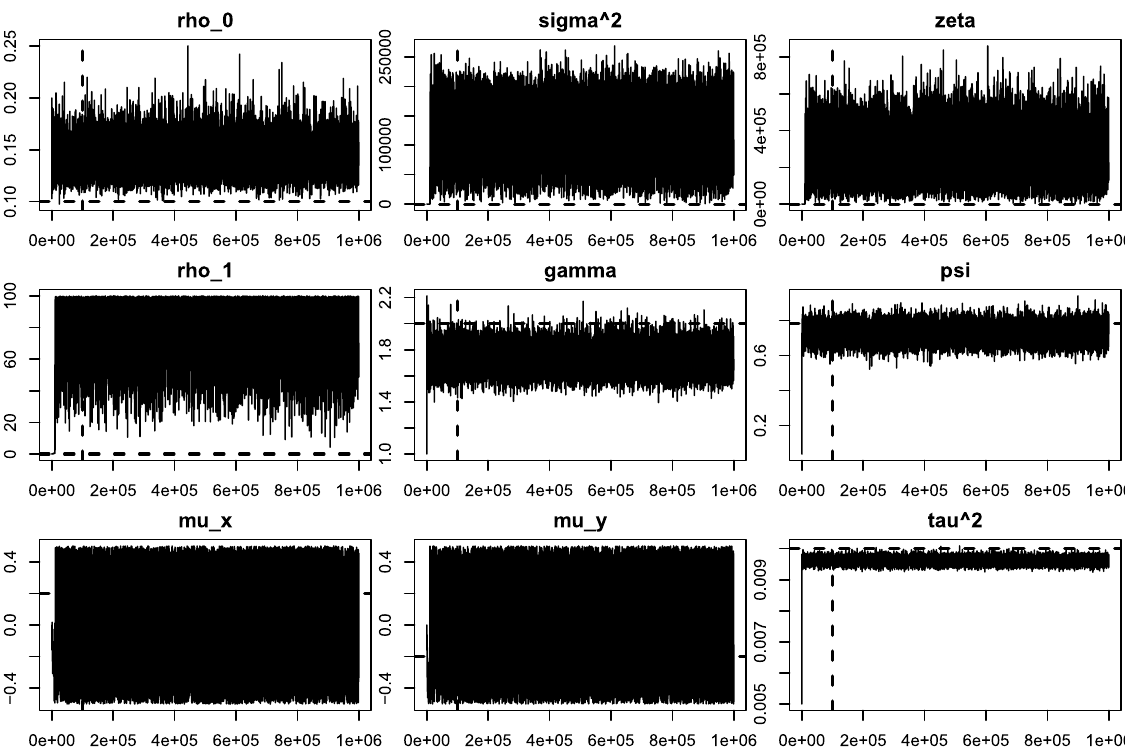}
				\caption{$\alpha=1.5$}
			\end{subfigure}
			
			&
			\begin{subfigure}{0.5\textwidth}
				\includegraphics[width=\textwidth]{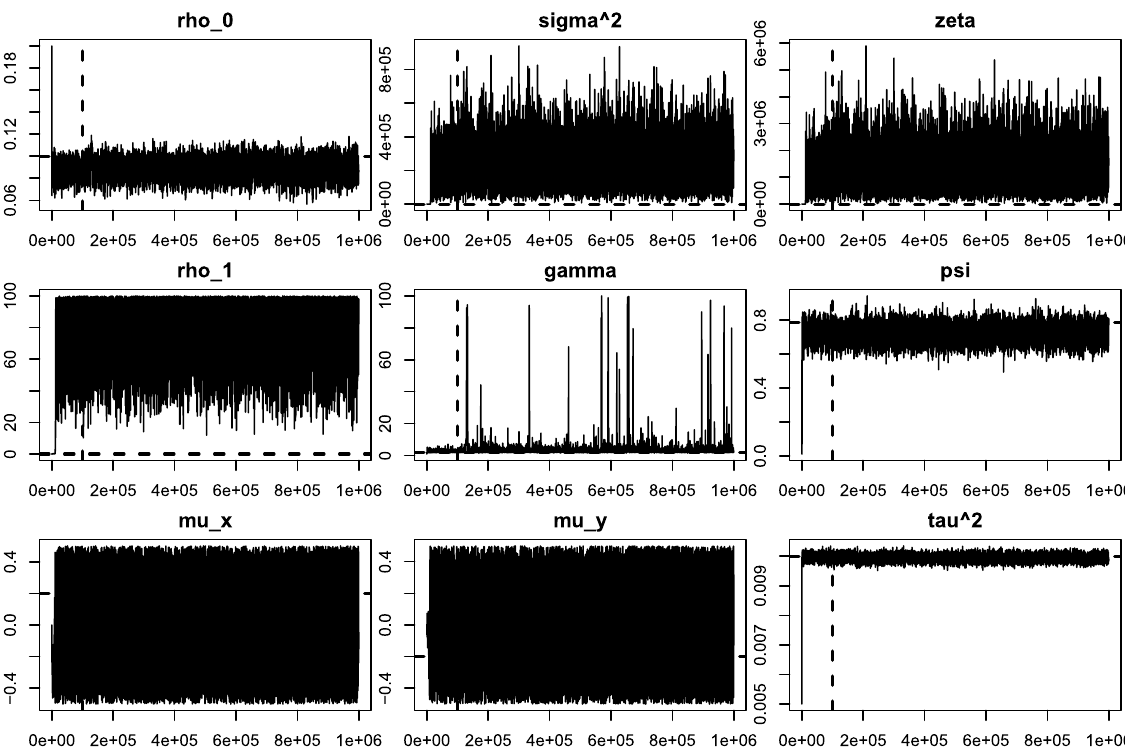}
				\caption{$\alpha=2.5$}
			\end{subfigure}
			
		\end{tabular}
		
		\caption{Trace plots of the parameters estimated when the algorithm knows the true $\alpha = 2$ value (top left), underestimates $\alpha = 1.5$ (bottom left), overestimates $\alpha = 2.5$ (bottom right) and estimates $\alpha$ (top right). The dashed line indicates the burn-in period. Assuming the wrong value of $\alpha$ leads to very poor parameter estimation with $\sigma^2, \zeta, \rho_1, \gamma, \mu_x, \mu_y, \tau$ in non-representative values. Samples for $\mu_x, \mu_y$ are still distributed according to their prior indicating no information was gained from the data for $\mu$. However estimating $\alpha$ and knowing the correct $\alpha$ allows for similar accurate parameter estimation results, as the samples are centered around or capture well the true value.}
		\label{fig:par_trace}
	\end{figure}
	
\end{document}